\documentclass[preprint]{tlp}

\usepackage{times}
\usepackage{latexsym}
\usepackage{amssymb}
\usepackage{amsmath}
\usepackage{xspace}
\usepackage{url}
\usepackage{graphicx}
\usepackage{subfigure}
\usepackage{epsfig}
\usepackage{listings}
\usepackage{lscape}
\usepackage{capt-of}

\newcommand{\DNF}{\ensuremath{\Delta S}\xspace}
\newcommand{\NNF}{\ensuremath{{\cal N}S}\xspace}

\newcommand{\nop}[1]{}
\newcommand{\hide}[1]{}
\newcommand{\mm}[1]{\mbox{${\rm MM}(#1)$}}

\newcommand{\gpii}{\mbox{$\Pi$}}
\newcommand{\NP}{{\rm NP}\xspace}

\newcommand{\hlb}[1][]{{\ensuremath{h_{LB}^{#1}}}\xspace}
\newcommand{\hlblf}[1][]{\hlb[LF]\xspace}
\newcommand{\hlbmf}[1][]{\hlb[MF]\xspace}
\newcommand{\hlblfre}[1][]{\hlb[LF,RE]\xspace}
\newcommand{\hlbmfre}[1][]{\hlb[MF,RE]\xspace}
\newcommand{\hlblfaf}[1][]{\hlb[LF,AF]\xspace}
\newcommand{\hlbmfaf}[1][]{\hlb[MF,AF]\xspace}
\newcommand{\hlblfreaf}[1][]{\hlb[LF,RE,AF]\xspace}
\newcommand{\hlbmfreaf}[1][]{\hlb[MF,RE,AF]\xspace}

\newcommand{\dlv}{{\sc DLV}\xspace}

\newcommand{\Or}{\ensuremath{\vee}}
\newcommand{\derives}{\ensuremath{\mbox{\,$:$--}\,}\xspace}

\newcommand{\naf}{\ensuremath{\mathtt{not\ }}}

\newcommand{\p}{\ensuremath{{\cal P}}\xspace}

\newcommand{\BP}{\ensuremath{B_{\p}}\xspace}
\newcommand{\UP}{\ensuremath{U_{\p}}\xspace}

\newenvironment{dlvcodel}
  {\begin{displaymath}\hspace{-0.24cm}\begin{array}{l}}
  {\end{array}\end{displaymath}}

\newcommand{\dgp}{\mbox{$G_{\p}$}}

\newenvironment{dlvcode}
  {\begin{displaymath}\begin{array}{l}}
  {\end{array}\end{displaymath}}

\newcommand{\serial}{\ensuremath{\mathtt{serial}}\xspace}
\newcommand{\kali}{\ensuremath{\mathtt{levels_{1\!+\!2}}}\xspace}
\newcommand{\splitonly}{\ensuremath{\mathtt{level_3}}\xspace}
\newcommand{\paral}{\ensuremath{\mathtt{levels_{1\!+\!2\!+\!3}}}\xspace}

\newif\ifmakebbl

\newcommand{\w}[1]{\ensuremath{{\cal W}(#1)}\xspace }

\title[Theory and Practice of Logic Programming]{ Parallel Instantiation of ASP Programs: \\ Techniques and Experiments}

   \author[Simona Perri, Francesco Ricca, Marco Sirianni]
{Simona Perri , Francesco Ricca,
Marco Sirianni\\
Dipartimento di Matematica, Universit{\`a} della Calabria, 87030 Rende, Italy\\
\email{\{perri,ricca,sirianni\}@mat.unical.it}}

 \submitted {28 November 2010}
 \revised {16 April 2011}
 \accepted{31 August 2011}

\begin{document}
\maketitle

\begin{abstract}

Answer Set Programming (ASP) is a powerful logic-based programming
language, which is enjoying increasing interest within the
scientific community and (very recently) in industry.
The evaluation of ASP programs is traditionally carried out in two
steps. At the first step an input program $\p$ undergoes the
so-called instantiation (or grounding) process, which produces a
program $\p'$ semantically equivalent to $\p$, but not containing
any variable; in turn, $\p'$ is evaluated by using a backtracking
search algorithm in the second step.
It is well-known that instantiation is important for the efficiency
of the whole evaluation, might become a bottleneck in common
situations, is crucial in several real-world applications, and is
particularly relevant when huge input data has to be dealt with.
At the time of this writing, the available instantiator modules are
not able to exploit satisfactorily the latest hardware, featuring
multi-core/multi-processor SMP (Symmetric MultiProcessing)
technologies.
This paper presents some parallel instantiation techniques, including
load-balancing and granularity control heuristics, which allow for the effective exploitation of
the processing power offered by modern SMP machines.
This is confirmed by an extensive experimental analysis herein reported.

{\em To appear in Theory and Practice of Logic Programming (TPLP)}.
\end{abstract}


\section{Introduction}\label{sec:introduction}
Answer Set Programming (ASP)~\cite{gelf-lifs-91,eite-etal-97f,lifs-99a,mare-trus-99,bara-2002,gelf-leon-02}
is a purely declarative programming paradigm based on nonmonotonic
reasoning and logic programming.
The language of ASP is based on rules, allowing (in general)
for both disjunction in rule heads and nonmonotonic negation in the body.
The idea of answer set programming is
to represent a given computational problem by a logic program such that its
answer sets correspond to solutions, and then, use an answer set solver
to find such solutions \cite{lifs-99a}.
The main advantage of ASP is its declarative nature combined with a
relatively high expressive power~\cite{leon-etal-2002-dlv,dant-etal-01}; but this comes at
the price of a elevate computational cost, which makes the
implementation of efficient ASP systems a difficult task.
Some effort has been made to this end, and, after some pioneering work
\cite{bell-etal-94,subr-etal-95}, there are nowadays a number of
systems that support ASP and its variants
\cite{leon-etal-2002-dlv,janh-etal-2005-tocl,simo-etal-2002,gebs-etal-2007-ijcai,lin-zhao-2004,lier-mara-2004-lpnmr,ange-etal-2001}. 
The availability of efficient systems made ASP profitably exploitable in real-world applications \cite{lemb-etal-02b,aiel-mass-2000,bara-uyan-2001}
and, recently, also in industry \cite{Grasso-etal-2010,ielpa-etal,Grasso-etal-2011}.

Traditionally, the kernel modules of such systems operate on a ground
instantiation of the input program, i.e. a program that does not
contain any variable, but is semantically equivalent to the
original input.
Therefore, an input program $\p$ first undergoes the so-called instantiation
process, which produces a ground program $\p'$ semantically equivalent to $\p$.
This phase, which is fundamental for some real world applications where huge amounts of input data have to be handled, is computationally expensive (see~\cite{eite-etal-97f,dant-etal-01});
and, nowadays, it is widely recognized that having an efficient instantiation procedure is
crucial for the performance of the entire ASP system.
Many optimization techniques have
been proposed for this purpose
\cite{fabe-etal-99c,leon-etal-2001a,perr-etal-2004-nmr};
nevertheless, the performance of instantiators can be further improved
in many cases, especially when the input data are
significantly large (real-world instances, for example, may count
hundreds of thousands of tuples).

In this scenario, significant performance improvements can be obtained
by exploiting modern (multi-core/multi-processor) Symmetric Multi Processing (SMP) \cite{stall-98} machines
featuring several CPUs.
In the past only expensive servers and workstations supported this technology, whereas,
at the time of this writing, most of the personal computers systems
and even laptops are equipped with (at least one) dual-core processor.
This means that the benefits of true parallel processing can
be enjoyed also in entry-level systems and PCs.
However, traditional ASP instantiators were not developed with
multi-processor/multi-core hardware in mind, and are unable to
exploit fully the computational power offered by modern machines.

This paper presents%
\footnote{Preliminary results have been presented in~\cite{perr-etal-2008,perr-etal-2010-damp}.}
 some advanced techniques for the parallel instantiation
of ASP programs, implemented in an instantiator system allowing the exploitation of the computational power
offered by multi-core/multi-processor machines.
The system is based on the state-of-the-art ASP instantiator
of the \dlv system~\cite{leon-etal-2002-dlv}; moreover, it
extends the work of~\cite{cali-etal-2008-joacil}
 by introducing a number of relevant improvements:
$(i)$ an additional third stage of parallelism for the instantiation of every single rule of the program,
$(ii)$ dynamic load balancing, and $(iii)$ granularity control strategies
based on computationally cheap heuristics.
In this way, the efficacy of the system is no longer limited to programs with
many rules (as in~\cite{cali-etal-2008-joacil}), and also  the particularly
(common and) difficult-to-parallelize class of programs with few rules
is handled in an effective way.
Moreover, we developed a new implementation supporting a richer input language (e.g. aggregates)
and technical improvements in thread management.

An extensive experimental activity is also reported, which was carried out on a variety
of publicly-available benchmarks already exploited for evaluating the performance of instantiation systems \cite{gebs-etal-2007-lpnmr-competition,devebogetr09a,leon-etal-2002-dlv}.
A comparison with~\cite{cali-etal-2008-joacil} shows that the new techniques
both combine with the previous ones and allow for a parallel evaluation
even in cases where previous techniques were not applicable.

A scalability analysis demonstrates that the new parallel instantiator behaves very well in all the considered
instances: superlinear speedups are observed in the case of easy-to-parallelize problem instances; and,
nearly optimal efficiencies are measured in the case of hard-to-parallelize problem instances
(its efficiency remains stable when the size of the input problem grows).
Importantly, the system offers a very good performance already when only two CPUs are enabled
(i.e. for the majority of the commercially available hardware at the time of this writing)
and efficiency remains at a good level when more CPUs are enabled.

The remainder of the paper is structured as follows:
Section~\ref{sec:dlp} outlines some basic notions of Answer Set
Programming;  Section~\ref{sec:parallel} describes
the employed parallel instantiation
strategies; Sections~\ref{sec:splitlit} and~\ref{sec:loadB} present heuristics, load balancing and granularity control;
Section~\ref{sec:experiments} discusses the results of
the experiments; finally, Section~\ref{sec:related} is devoted to
related works, and Section~\ref{sec:conclusion} draws some
conclusions.

\section{Answer Set Programming}\label{sec:preliminaries}\label{sec:dlp}\label{sec:ASP}

In this section, we provide a formal definition of syntax and semantics of ASP programs.

\noindent {\em \bf Syntax.}
A variable or a constant is a {\em term}.  An {\em atom} is
$p(t_{1},..., t_{n})$, where $p$ is a {\em predicate} of arity $n$
and $t_{1},..., t_{n}$ are terms.  A {\em literal} is either a {\em
positive~literal} $p$ or a {\em negative~literal} $\naf p$, where
$p$ is an atom.
 A {\em (disjunctive) rule} $r$ has the following
form:
\begin{dlvcodel}
a_1\lor\ldots\lor a_n \derives
        b_1,\ldots, b_k,
        \naf b_{k+1},\ldots, \naf b_m.  \quad \quad n\geq 0,\ m\geq k\geq 0
\end{dlvcodel}
\hspace{-0.25cm} where $a_1,\ldots ,a_n,b_1,\ldots ,b_m$ are atoms.
 The
disjunction $a_1\lor\ldots\lor a_n$ is the {\em head} of $r$, while
the conjunction $b_1 , \ldots,  b_k, \naf b_{k+1} , \ldots, \naf
b_m$ is the {\em body} of $r$.

We denote by $H(r)$ the set $\{a_1, \ldots, a_n \}$ of the head
atoms, and by $B(r)$ the set $\{b_1, \ldots,$ $b_k,$ $\naf b_{k+1}
,\ldots , \naf b_m \}$ of the body literals. $B^+(r)$ (resp.,
$B^-(r)$) denotes the set of atoms occurring positively (resp.,
negatively) in $B(r)$.

A rule having precisely one head literal (i.e.\ $n=1$) is called a
{\em normal rule}. If the body is empty (i.e.\ $k=m=0$), it is
called a {\em fact}, and we usually omit the $\derives$ sign. A rule
without head literals (i.e.\ $n=0$) is usually referred to as an
{\em integrity constraint}.\footnote{Note that a constraint is a
shorthand for ${\mathit false} \derives b_1,\ldots, b_k,\ \naf\
b_{k+1},\ldots,\ \naf\ b_m.$, and it is also assumed that a rule
${\mathit bad} \derives {\mathit false}, \naf {\mathit bad}$ is in
the program, where $\mathit{ false}$ and $\mathit{ bad}$ are special
symbols appearing nowhere else in the original program. } A rule $r$
is {\em safe} if each variable appearing in $r$ appears also in some
positive body literal of $r$.

An {\em ASP program} $\p$ is a finite set of safe rules. A
$\naf$-free (resp., $\lor$-free) program is called {\em positive}\/
(resp., {\em normal}).  A term, an atom, a literal, a rule, or a
program is {\em ground} if no variables appear in it.
A predicate $p$ is referred to as an {\em EDB} predicate if, for
each rule $r$ having in the head an atom whose name is $p \in H(r)$, $r$ is a fact; all others
predicates are referred to as {\em IDB} predicates. The set of facts
in which {\em EDB} predicates occur, denoted by $EDB(\p)$, is called {\em Extensional
Database (EDB)}, the set of all other rules is the {\em Intensional Database (IDB)}.

\noindent {\em \bf  Semantics.}
Let $\p$ be an ASP program. The {\em Herbrand universe}
of $\p$, denoted as $U_{\p}$, is the set of all constants appearing
in $\p$. In the case when no constant appears in $\p$, an arbitrary constant
$\psi$ is added to $U_{\p}$. The {\em Herbrand base} of \p{},
denoted as $B _{\p}$, is the set of all ground atoms constructible
from the predicate symbols appearing in $\p$ and the constants of
$U_{\p}$.
Given a rule $r$ occurring in a program $\p$, a {\em ground
instance} of  $r$ is a rule obtained from $r$ by replacing every
variable $X$ in $r$ by $\sigma (X)$, where $\sigma$
 is a substitution mapping the variables occurring in $r$ to
constants in $\UP$. We denote by $ground( \p)$ the set of all the
ground instances of the rules occurring in $\p$.

 An {\em interpretation} for $\p$ is a set of ground atoms, that
is, an interpretation is a subset $I$ of $\BP$. A ground positive
literal $A$ is true (resp., false) w.r.t. $I$ if
 $A \in I$ (resp.,  $A \not\in I$). A ground
negative literal $\naf A$ is true w.r.t. $I$ if $A$ is false
w.r.t. $I$; otherwise $\naf A$ is false w.r.t. $I$.
Let  $r$ be a rule in $ground( \p )$.  The head of $r$ is
true w.r.t. $I$ if $H(r) \cap I \neq \emptyset$. The body of
 $r$ is true w.r.t. $I$ if all body
literals of  $r$ are true w.r.t.  $I$ (i.e.,  $B^+(r) \subseteq I$
and  $B^-(r)\cap I = \emptyset$) and otherwise the body of $r$ is false w.r.t. $I$.
The rule  $r$ is {\em satisfied} (or true) w.r.t.
 $I$ if its head is true w.r.t.  $I$ or its body
is false w.r.t.  $I$.
A {\em model} for  $\p$
 is an interpretation  $M$ for
 $\p$ such that every rule  $r \in ground(\p)$ is
true w.r.t.  $M$.  A model  $M$ for  $\p$ is {\em minimal} if there is no
model  $N$ for $\p$
such that $N$ is a proper subset of  $M$. The set of all minimal
models for  $\p$ is denoted by  ${\rm MM}(\p )$.

In the following, the semantics of ground programs is first given,
then the semantics of general programs is given in terms of the answer sets of its instantiation.

Given a {\em ground} program  $\p$ and an interpretation  $I$, the
{\em reduct} of $\p$ w.r.t. $I$ is the subset $\p^I$ of $\p$ obtained
by deleting from $\p$ the rules in which a body literal is false w.r.t. $I$.

Note that the above definition of reduct, proposed in
\cite{fabe-etal-2004-jelia}, simplifies the original definition of
Gelfond-Lifschitz (GL) transform~\cite{gelf-lifs-91}, but is fully
equivalent to the GL transform for the definition of answer sets
\cite{fabe-etal-2004-jelia}.

Let  $I$ be an interpretation for a ground program  $\p$.  $I$ is an {\em answer set} (or stable model)
for  $\p$ if $I \in \mm{\p^I}$ (i.e.,  $I$ is a minimal model for the program $\p^I$) \cite{fabe-etal-2004-jelia}.
The set of all answer sets for $\p$ is denoted by $ANS(\p)$.

\section{Parallel Instantiation}\label{sec:parallel}

In this section we describe an algorithm for the instantiation
of ASP programs, which exploits parallelism in three different
steps of the instantiation process. In particular, the algorithm
employs techniques presented in~\cite{cali-etal-2008-joacil}
and integrates them with a novel strategy which has a
larger application field, covering
many situations in which the previous techniques do not apply.
More in detail, the parallel instantiation algorithm
allows for three levels of parallelism:  {\em components}, {\em rules} and
{\em single rule} level. The first level allows for instantiating in
parallel subprograms of the program in input: it is especially
useful when handling programs containing parts that are,
somehow, independent. The second one allows for
the parallel evaluation of rules within a given subprogram: it is
 useful when the number of rules in the subprograms is large. The
third new one, allows for the parallel evaluation of a single rule:
it is crucial for the parallelization of
programs with few rules, where the first two levels are almost not
applicable.
In the following, we first provide an overview of our approach to the
parallel instantiation process, giving an intuition of the three
aforementioned levels and then we illustrate
the instantiation algorithm.

\subsection{Overview of the Approach}\label{subsec:paroverview}
\paragraph {Components Level}\cite{cali-etal-2008-joacil}.
The first level of parallelism, called {\em Components Level},
consists in dividing the input program \p into subprograms,
according to the dependencies among the IDB predicates of \p, and by
identifying which of them can be evaluated in parallel.
More in detail, each program \p is associated with a graph, called
the {\em Dependency Graph} of \p, which, intuitively, describes how
IDB predicates of \p depend on each other. For a program \p, the
{\em Dependency Graph} of \p is a directed graph $\dgp = \left
\langle N, E \right \rangle $, where $N$ is a set of nodes and $E$
is a set of arcs. $N$ contains a node for each IDB predicate of \p,
and $E$ contains an arc $e = (p, q)$ if there is a rule $r$ in $\p$
such that $q$ occurs in the head of $r$ and $p$ occurs in a positive
literal of the body of $r$.

The graph \dgp\ induces a subdivision of \p into subprograms (also
called {\em modules}) allowing for a modular evaluation. We say that
a rule $r\in \p$ {\em defines} a predicate $p$ if $p$ appears in the
head of $r$. For each strongly connected component%
\footnote{A
strongly connected component of a directed graph is a maximal subset
of the vertices, such that every vertex is reachable from every
other vertex.}
(SCC) $C$ of \dgp , the set of rules defining all the
predicates in $C$ is called {\em module} of $C$.
Moreover, a partial ordering among the SCCs is induced by \dgp,
defined as follows: for any pair of SCCs $A$, $B$ of \dgp , we say
that $B$ directly depends on $A$ if there is an arc from a predicate
of $A$ to a predicate of $B$; and, $B$ depends on $A$ if there is a
path in \dgp\ from $A$ to $B$.

According to such definitions, the instantiation of the input
program \p can be carried out by separately evaluating its modules;
if the evaluation order of the modules respects the above mentioned
partial ordering then a small ground program is produced \cite{cali-etal-2008-joacil}. Indeed,
this gives the possibility of computing ground instances of rules
containing only atoms that can possibly be derived from \p
(thus, avoiding the combinatorial explosion that can be obtained by
naively considering all the atoms in the Herbrand base).

Intuitively, this partial ordering guarantees that a component $A$
precedes a component $B$ if the program module corresponding to $A$
has to be evaluated before the one of $B$, because the evaluation of
A produces data that are needed for the instantiation of B.
Moreover, the partial ordering allows for determining which modules
can be evaluated in parallel. Indeed, if two components $A$ and $B$,
do not depend on each other, then the instantiation of the
corresponding program modules can be performed simultaneously,
because the instantiation of $A$ does not require the data produced
by the instantiation of $B$ and vice versa.
The dependency among components is thus the principle underlying the
first level of parallelism. At this level subprograms can be
evaluated in parallel, but still the evaluation of each subprogram
can be further parallelized.

\paragraph{Rules Level}\cite{cali-etal-2008-joacil}.
The second level of parallelism, called the {\em Rules Level},
allows for concurrently evaluating the rules within each module. A
rule $r$ occurring in the module of a component $C$ (i.e.,
defining some predicate in $C$) is said to be {\em recursive} if
there is a predicate $p \in C$ occurring in the positive body of
$r$; otherwise, $r$ is said to be an {\em exit rule}. Rules are
evaluated following a semi-na\"ive schema~\cite{ullm-89} and the
parallelism is exploited for the evaluation of both exit and
recursive rules.
More in detail, for the instantiation of a module $M$, first all
exit rules are processed in parallel by exploiting the data (ground
atoms) computed during the instantiation of the modules which $M$
depends on (according to the partial ordering induced by the
dependency graph).
Only afterward, recursive rules are processed in parallel several
times by applying a semi-na\"ive evaluation technique in which, at
each iteration $n$, the instantiation of all the recursive rules is
performed concurrently and by exploiting only the significant
information derived during iteration $n-1$.
\nop{ This is done by
partitioning significant atoms into three sets: $\Delta S$, $S$ and
$NS$. $NS$ is filled with atoms computed during current iteration
(say $n$); $\Delta S$ contains atoms computed during previous
iteration (say $n-1$); and, $S$ contains the ones previously
computed (up to iteration $n-2$).

Initially, $\Delta S$ and $NS$ are empty; while $S$ contains all the
information previously derived in the instantiation process. At the
beginning of each new iteration, $NS$ is assigned to $\Delta S$,
i.e. the new information derived during iteration $n$ is considered
as significant information for iteration $n+1$. Then, the recursive
rules are processed simultaneously and each of them uses the
information contained in the set $\Delta S$; at the end of the
iteration, when the evaluation of all rules is terminated, the set
$\Delta S$ is added to the set $S$ (since it has already been
exploited).  The evaluation stops whenever no new information has
been derived (i.e. $NS=\emptyset$).
}
\paragraph{Single Rule Level.}\label{sec:rew}
The first two levels of parallelism are effective when handling large programs.
However, when the input program consists of few rules, their efficacy is drastically reduced, and
there are cases where components and rules parallelism are not
exploitable at all. For instance the following program \p encoding the
well-known 3-colorability problem:

\begin{small}
\begin{dlvcode}
(r)\ \ \ col(X,red)\ \Or\ col(X,yellow)\ \Or\ col(X,green)\ \derives\ node(X). \\
(c)\ \ \ \derives\ edge(X,Y), col(X,C),\ col(Y,C).
\end{dlvcode}
\end{small}

\noindent The two levels of parallelism described above have no
effects on the evaluation of \p. Indeed, this encoding consists of
only two rules which have to be evaluated sequentially, since,
intuitively, the instantiation of $(r)$ produces the ground atoms
with predicate $col$, which are necessary for the evaluation of
$(c)$.

For the instantiation of this kind of programs a third level is
necessary for the parallel evaluation of each single rule, which is
therefore called {\em Single Rule Level}.

In the following we present a strategy  for parallelizing the evaluation of a rule. The idea is to
partition the extension of a single rule literal (hereafter called {\em split} literal)
 into a number of subsets. Thus  the rule instantiation is divided into a number of smaller similar tasks
each of which considers as extension of the split literal only one of those subsets.
For instance, the evaluation of rule $(c)$ in the previous example can be performed in parallel by
partitioning the extension of one of its literals, let it be $edge$, into $n$ subsets, thus obtaining
$n$ instantiation tasks for $(c)$, working with different ground instances of $edge$.
Note that, in general, several body literals are possible candidates to be split up
(for instance, in the case of $(c)$, $col$ can be split up instead
of $edge$) and the choice of the most suitable literal to split has to be
carefully made, since it may strongly affect the cost of the
instantiation of rules. Indeed, a ``bad'' split might reduce or
neutralize the benefits of parallelism, thus making the overall time
consumed by the parallel evaluation not optimal (and, in some corner
cases, even worse than the time required to instantiate the original
encoding).
Note also that, the partitioning of the extension of the split literal has to be performed at run-time.
Indeed, if  the predicate to split is an IDB predicate,
as in the case of $col$, the partitioning can be made only when the extension of the IDB predicate has
already been computed; in our example, only after the evaluation of rule $(r)$.%
\footnote{An example of workload distribution is reported in Appendix C.}

\begin{figure}[t!]
\begin{center}
\begin{small}
\begin{tabbing}

// {\em First level of parallelism}\\
{\bf Procedure} $\mathit{Components\_Instantiator \/}$(\= $\p$: Program;
$\dgp$: DependencyGraph; {\bf var} $\gpii$: GroundProgram)\\;
\hspace*{0.2cm} \= \kill
\hspace*{0.4cm} \= \hspace*{0.4cm} \= \hspace*{0.4cm} \=
\hspace*{0.4cm} \= \hspace*{0.4cm} \= \hspace*{3cm} \=\kill
{\bf begin} \\
\> {\bf var} $S$: SetOfAtoms; {\bf var} $C$: SetOfPredicates;\\
\> $S = EDB(\p)$; $\gpii := \emptyset$; \\
\> {\bf while} $\dgp \neq \emptyset$ {\bf do}  \hspace{0.7 cm}{\em // until there are components to be processed} \\
\>\> take a SCC $C$ from $\dgp$ that does not depend on other SCCs  of $\dgp$    \\
\>\> {\em Spawn}({\em Rules\_Instantiator},\ $\p,C,S,\gpii,$\dgp$)$\\
\> {\bf end while} \\
{\bf end}; \\

\\

// {\em Second level of parallelism}\\
{\bf Procedure} $\mathit{Rules\_Instantiator\/}$\=($\p$: Program; $C$: Component;
                                     {\bf var} $S$: SetOfAtoms; \\
                                     \> {\bf var} $\gpii$: GroundProgram, {\bf var} $\dgp$: DependencyGraph)\\
\hspace*{0.2cm} \= \kill
\hspace*{0.4cm} \= \hspace*{0.4cm} \= \hspace*{0.4cm} \=
\hspace*{0.4cm} \= \hspace*{0.4cm} \= \hspace*{3cm} \=\kill
{\bf begin} \\
\> {\bf var} $\DNF$, $\NNF$: SetOfAtoms;\\
\> $\DNF := \emptyset$; $\NNF := \emptyset$ ; \\
\> {\bf for each} $r \in Exit(C, \p)$ {\bf do} \hspace{1.5cm} {\em // evaluation of exit rules} \\
\>\>$\mathcal{I}_r$ = {\em Spawn} $(SingleRule\_Instantiator, r, S,\DNF,\NNF,\gpii)$; \\
\> {\bf for each} $r \in Exit(C, \p)$ {\bf do} \hspace{1.48cm} {\em // synchronization barrier} \\
\>\> {\em join\_with\_thread}($\mathcal{I}_r$); \\
\>{\bf do} \\
\>\> $\DNF := \NNF$; $\NNF := \emptyset$ ; \\
\>\>  {\bf for each} $r \in Recursive(C, \p)$ {\bf do}  \hspace{0.2cm} {\em // evaluation of recursive rules} \\
\> \> \> $\mathcal{I}_r$ = {\em Spawn} $(SingleRule\_Instantiator, r, S,\DNF,\NNF,\gpii)$; \\
\>\>  {\bf for each} $r \in Recursive(C, \p)$ {\bf do} \hspace{0.2cm} {\em // synchronization barrier} \\
\>\>\> {\em join\_with\_thread}($\mathcal{I}_r$); \\
\> \> $S := S \cup \DNF$; \\
\> {\bf while \ \ $\NNF \neq \emptyset$} \hspace{3.2cm} {\em // until no new information can be derived} \\
\> Remove $C$ from $\dgp$; \hspace{2.55cm} {\em // to process $C$ only once} \\
{\bf end Procedure}; \\ \\
\end{tabbing}
\end{small}
\caption{Components and Rules parallelism}\label{fig:parallelInstantiation1}
\end{center}
\end{figure}

\begin{figure}[th!]
\begin{center}
\begin{small}
\begin{tabbing}

// {\em Third level of parallelism}\\ \
\\
// {\em A split is virtually identified by four iterators to $S$ and $\DNF$ identifying ranges of instances. }\\
struct VirtualSplit
\{
AtomsIterator $S\_begin$, $S\_end$, $\DNF\_begin$, $\DNF\_end$;
\}\\ \
\\
{\bf Procedure} $\mathit{  SingleRule\_Instantiator\/}$\=($r$: Rule; $S$: SetOfAtoms; \DNF: SetOfAtoms; \NNF: SetOfAtoms;\\
                                     \> {\bf var} $\gpii$: GroundProgram)\\
\hspace*{0.2cm} \= \kill
\hspace*{0.4cm} \= \hspace*{0.4cm} \= \hspace*{0.4cm} \=
\hspace*{0.4cm} \= \hspace*{0.4cm} \= \hspace*{3cm} \=\kill
\> {\bf var} $s:= \text{numberOfSplits}(B(r),S,\DNF)$; // {\em  according to load balancing and granularity control} \\
\> SelectLiteralToSplit($L$,$B(r)$,$s$); \hspace{0.82cm} // {\em  according to a heuristics} \\
\> {\bf var} vector$<$VirtualSplit$>$ $Splits[s]$; \ \  // {\em  create virtual splits} \\
\> SplitExtension($L$, $s$, $S$, \DNF, $Splits$); \hspace{0.7 cm} // {\em  distribute extension of $L$} \\
\> {\bf for each} $sp$ in $Splits$ \hspace{2.05 cm}// {\em  spawn threads running InstantiateRule for each split} \\
\>\> $\mathcal{I}_{sp}$ = {\em Spawn} $(InstantiateRule, r, L, sp, S,\DNF,\NNF,\gpii)$; \\
\> {\bf for each} $sp$ in $Splits$ {\bf do} \hspace{1.5 cm} // {\em synchronization barrier} \\
\>\> {\em join\_with\_thread}($\mathcal{I}_{sp}$); \\
{\bf end Procedure}; \\

\\

{\bf Procedure} $\mathit{ SplitExtension \/}$(\= $L$: Literal; $s$:integer; $S$: SetOfAtoms; $\DNF$: SetOfAtoms; \\
\> {\bf var} vector$<$VirtualSplit$>$ $Splits$)\\
\hspace*{0.2cm} \= \kill
{\bf /*}
\>  {\em Given a literal $L$ to split, a number $s$ of splits to be produced, virtually partitions the extension}\\
\> { \em  of $L$ stored in $S$ and $\DNF$ by determining a vector $Splits$ of VirtualSplit structures identifying} \\
\>  {\em  ranges of instances to be considered in each split. }\\
{\bf */}
\\ \
\\
{\bf Procedure} $\mathit{ InstantiateRule \/}$(\= $r$: rule; $L$: Literal; $sp$: VirtualSplit; $S$: SetOfAtoms; $\DNF$: SetOfAtoms;\\
\>{\bf var} $\NNF$: SetOfAtoms; {\bf var} $\gpii$: GroundProgram)\\
\hspace*{0.2cm} \= \kill
{\bf /*}
\  {\em Given $S$ and $\DNF$ builds all the ground instances of $r$, adds them to $\Pi$, and add to $\NNF$ } \\
\> \ {\em  the new head atoms of the generated ground rules. For $L$ only the ground atoms belonging to }\\
\> \ {\em the ranges \{S\_Begin,S\_End\} and \{\DNF\_begin,\DNF\_end\} indicated by $sp$ are used  .}\\
{\bf */}
\end{tabbing}

\end{small}
\caption{Single Rule  parallelism}\label{fig:parallelInstantiation2}
\end{center}
\end{figure}

\subsection{The Algorithms}

The algorithms for the three levels of parallelism mentioned above are
shown in Fig.~\ref{fig:parallelInstantiation1}, and Fig.~\ref{fig:parallelInstantiation2}.
They repeatedly apply a pattern similar to the classical producer-consumers problem. A
\textit{manager} thread (acting as a producer) identifies the
tasks that can be performed in parallel  and delegates their instantiation to a number
of \textit{worker} threads (acting as consumers).
More in detail, the {\em Components\_Instantiator} procedure (see Fig.~\ref{fig:parallelInstantiation1}),  acting as a manager,
implements the first level of parallelism, that is the parallel evaluation of program modules.
It receives as input both a
program $\p$ to be instantiated and its Dependency Graph $\dgp$; and it
outputs a set of ground rules $\Pi$, such that $ANS(\p)=ANS( \gpii
\cup EDB(\p))$.
First of all, the algorithm creates a new set $S$ of atoms that will
contain the subset of the Herbrand base significant for the
instantiation; more in detail, $S$ will contain, for each predicate $p$ in the program,
the extension of $p$, that is, the set of all the ground atoms having the predicate name of $p$
 (significant for the instantiation).

Initially, $S=EDB(\p)$, and $\Pi = \emptyset$. Then,
the manager checks whether some SCC $C$ can be instantiated; in particular,
it checks whether there is some other component $C'$ that has not been evaluated yet and such that $C$ depends on $C'$.
As soon as a component $C$ is processable,  a new thread is created, by a call to threading function {\em Spawn}, running procedure
{\em Rules\_Instantiator}.

Procedure {\em Rules\_Instantiator} (see Fig.~\ref{fig:parallelInstantiation1}), implementing the second level of parallelism,
takes as input, among the others, the component $C$ to be instantiated and the set $S$;
for each atom $a$ belonging to $C$, and for each rule $r$ defining
$a$, it computes the ground instances of $r$ containing only atoms
that can possibly be derived from $\p$. At the same time, it
updates the set $S$ with the atoms occurring in the heads of the
rules of $\Pi$. To this end, each rule $r$ in the program module of
$C$ is processed by calling procedure {\em
SingleRule\_Instantiator}.

It is worth noting that exit rules are instantiated by a single
call to {\em SingleRule\_Instantiator}, whereas recursive ones are
processed several times according to a semi-na\"ive evaluation
technique~\cite{ullm-89}, where at each iteration $n$ only the
significant information derived  during iteration $n-1$ is
used. This is implemented by partitioning significant atoms into
three sets: \DNF, $S$, and \NNF. \NNF is filled with atoms computed
during current iteration (say $n$); \DNF contains atoms computed
during previous iteration (say $n-1$); and, $S$ contains the ones
previously computed (up to iteration $n-2$).

 Initially, \DNF and \NNF are empty; the exit rules contained in the program module
of $C$ are evaluated  and, in particular, one new thread identified by $\mathcal{I}_r$ for each
exit rule $r$, running procedure {\em SingleRule\_Instantiator}, is
spawned. Since the evaluation of recursive rules has to be performed only
when the instantiation of all the exit rules is completed, a synchronization barrier is
exploited. This barrier is encoded (\`{a} la POSIX) by several calls to threading function {\em join\_with\_thread}
forcing {\em Rules\_Instantiator} to wait until all {\em SingleRule\_Instantiator} threads are done.
Then, recursive rules are
processed (do-while loop). At the beginning of each iteration, \NNF
is assigned to \DNF, i.e. the new information derived during
iteration $n$ is considered as significant information for iteration
$n+1$. Then, for each recursive rule, a new thread is spawned,
running procedure {\em SingleRule\_Instantiator}, which receives as
input $S$ and $\DNF$; when all threads terminate (second barrier), \DNF is added to
$S$ (since it has already been exploited). The evaluation stops
whenever no new information has been derived (i.e. $\NNF =
\emptyset$). Eventually, $C$ is removed from $\dgp$.

The third level of parallelism (see Fig.~\ref{fig:parallelInstantiation2}), concerning the parallel evaluation
of each single rule, is then implemented by Procedure {\em SingleRule\_Instantiator},
which given the sets $S$ and $\DNF$ of  atoms that are known to be significant up to now, builds all the
ground instances of $r$ and adds them to $\Pi$.
Initially, {\em SingleRule\_Instantiator} selects, according to a
heuristics for load balancing (see Section \ref{sec:loadB}) the
number $s$ of parts in which the evaluation has to be divided; then
{\em SingleRule\_Instantiator} heuristically selects a positive
literal to split in the body of $r$, say $L$ (see Section
\ref{sec:splitlit}). A call to function {\em SplitExtension}
(detailed in Appendix A) partitions the extension of $L$ (stored in
$S$ and $\DNF$) into $s$  equally sized parts, called splits. In
order to avoid useless copies, each split is virtually identified by
means of iterators over $S$ and $\DNF$, representing ranges of
instances. More in detail, for each  split, a VirtualSplit is
created containing two iterators over $S$ (resp. \DNF), namely
$S\_begin$ and $S\_end$ (resp. $\DNF\_begin$ and $\DNF\_end$),
indicating the instances of $L$ from $S$ (resp. \DNF) that belong to
the split. Note that, in general, a split may contain ground atoms
from both $S$ and \DNF. Once the extension of the split literal has
been partitioned, then a number of threads running procedure {\em
InstantiateRule}, are spawned. {\em InstantiateRule}, given $S$ and
\DNF builds all the ground instances of $r$ that can be obtained by
considering as extension of the split literal $L$ only the ground
atoms indicated by the iterators in the virtual split at hand. {\em
SingleRule\_Instantiator} terminates (last barrier) once all splits
are evaluated.

The correctness of the algorithm follows from the consideration
that, whatever the split literal $L$, the union of the outputs of
all the $s$ concurrent {\em InstantiateRule} procedures is the same
as the output produced by a single call to {\em InstantiateRule}
working with the entire extension of $L$ ($s=1$). Note that, if the
split predicate is recursive, its extension may change at each
iteration. This is considered in our approach by performing
different splits of recursive rules at each iteration. This ensures
that at each iteration the virtual splits are updated according to
the actual extension of the literal to split.

In addition, this choice has a relevant side-effect: at each
iteration the workload is dynamically re-distributed among
instantiators, thus inducing a form of dynamic load balancing
in case of the evaluation of recursive rules.

\section{Selection of the Literal to Split}\label{sec:splitlit}

Concerning the selection of the literal to split, the choice has to
be carefully made, since it may strongly affect the cost of the
instantiation of rules. It is well-known that this cost strongly depends on
the order of evaluation of body literals, since computing all the possible
instantiations of a rule is equivalent to computing all the answers
of a conjunctive query joining the extensions of literals of the
rule body. However, the choice of the split literal may
influence the time spent on instantiating each split rule, whatever the join order.
In the light of these observations, we have devised a new heuristics for selecting
the split literal {\em given an optimal ordering} (which can be obtained as explained in \cite{leon-etal-2001a}).

Intuitively, suppose we have a rule $r$ containing $n$ literals in
the body in a given order, and suppose that any body literal allows
for the target number of splits, say $s$, then: to obtain work for
$s$ threads it is sufficient to split the first literal (whatever
the join order); nonetheless, moving forward, say splitting the
third literal, would cause a replicate evaluation of the join of the
first two literals in each split thread possibly increasing parallel
time. It is easy to see that the picture changes if  all/some body
literal does not allow for the target number of splits, in this case
one should estimate the cost of splitting a literal different from
the first and select the best possible choice.

In the following, we first introduce some metrics for estimating the
work needed for instantiating a given rule, and then we describe the
new heuristics.
In detail, we use the following estimation for determining
the size of the joins of the body literals: given two relations $R$ and $S$,
with one or more common variables, the size of $R \Join S$ can be estimated
as follows:
\begin{equation}
\label{eq_1}
T\left(R \Join S \right) = \frac{T\left( R\right) \cdotp T\left( S\right)} {\prod _{X \in var\left(R \right) \cap var\left(S \right)} \max \left\lbrace V\left(X,R \right),V\left(X,S\right)\right\rbrace }
\end{equation}

\noindent where $T\left(R \right) $ is the number of tuples in $R$,
and $V\left(X,R \right) $ (called selectivity) is the number of distinct values
assumed by the variable $X$ in $R$. Given an evaluation order of body literals,
one can repeatedly apply this formula to pairs of body predicates
for estimating the size of the join of a body.
A more detailed discussion on this estimation can be found in~\cite{ullm-89}.

Let $r$ be a rule with $n$ body literals $L_1,L_2, \dots ,L_n$,
where $L_i$ precedes $L_j$ for each $i<j$ in a given evaluation order,
an estimation of the cost of instantiating the first $k$ literals in $B(r)$ is:
\begin{equation}
\label{eq_2}
{\cal C}(k) \ = \
\left\{
\begin{array}{@{}l@{}}
0 \ \ \ if \ k < 2\\
T(L_1) \cdot T(L_2) \ \ \ if \ k = 2\\
{\cal C}(k-1) + T(L_1 \Join \dots \Join L_{k-1}) \cdot T(k)\ \ \ if \ k>2\\
\end{array}
\ \right.
\end{equation}

Now, let $s$ be the number of splits to be performed; the following is
an estimation of the work of the instantiation tasks obtained by the split of the $i$-th literal $L_i$:
\begin{equation}
\label{eq_3}
{\cal C}^i \ = \ \frac{{\cal C}(n)-{\cal C}(i-1)}{s^i} +{\cal C}(i-1), \ \ 1\leq i \leq n
\end{equation}
where, $s^i$ is  equal to $s$ (if the extension of $L_i$ is sufficiently large) or the maximum number of splits allowed by $L_i$.
Intuitively, if $L_i$ is the split literal, the work of each instantiation task is composed of two parts:
a part to be performed serially, common to all tasks, which consists in the instantiation of the first $i-1$ literals,
whose cost is represented by ${\cal C}(i-1)$;
and the instantiation of the remaining literals, which is divided among the $s^i$ tasks, whose cost is represented by
$\frac{{\cal C}(n)-{\cal C}(i-1)}{s^i}$.
The estimation ${\cal C}^i$ can be used for determining the split literal, by choosing the one with minimum cost.

Note that, in the search for the best one, we can skip over each body literal $L_k$, with $k>j$, if $L_j$ allows for $s$ splits since ${\cal C}^j$ $\leq$  ${\cal C}^k$ holds.
Indeed, if $n = 2$, ${\cal C}^1 \ = \ {\cal C}^2$; while for $n \geq 3$,  $k = j+1$ and $s^j = s^k = s$ (worst case) we have that
$${\cal C}^k \ = \ \frac{{\cal C}(n)-{\cal C}(k-1)}{s^k} +{\cal C}(k-1) = \ \frac{{\cal C}(n)-{\cal C}(j)}{s} +{\cal C}(j).$$
By applying~(\ref{eq_2}):
$$ {\cal C}^k = \ \frac{{\cal C}(n)-{\cal C}(j-1)}{s}-\frac{Q}{s} +{\cal C}(j-1) + Q = C^j+ \frac{s-1}{s} \cdot Q$$ where $Q = T(L_1 \Join \dots \Join L_{j-1}) \cdot T(j)$.
Thus, by induction, if $L_j$ allows for $s$ splits, ${\cal C}^j$ $\leq$  ${\cal C}^k$, for $k>j$.
Intuitively, this can be explained by considering that splitting a literal $L_k$ after one allowing for $s$ splits $L_j$
has the effect of evaluating serially the join of literals between $L_k$ and $L_j$ thus leading to a greater evaluation time.
Clearly, even a literal $L$ whose extension cannot be split in $s$ parts can be chosen, provided that $L$ allows for a minor (estimated) work for each instantiation task.
Moreover, if $s^1 = s$ ($L_1$ allows for $s$ splits), it holds that ${\cal C}^1 \leq  {\cal C}^i$, for each $1< i\leq n$;
in this case, $L_1$ can be chosen without computing any cost.

As an example, suppose that we have to instantiate the constraint
$\derives a(X,Y), b(Y,Z),$ $c(Z,X), d(V,Z).$
Suppose also that the extensions of the body literals are $T(a)=20$, $T(b)=50$, $T(c)=1000$, $T(d)=1000$,
and that the estimations of the costs of instantiating the first $i$ literals with $1\leq i\leq 4$ are the following:
${\cal C}(1)= 0$, ${\cal C}(2)=1000$, ${\cal C}(3)=7000$, ${\cal C}(4)=57000$.
Table~\ref{funValues} shows the estimations ${\cal C}^i$ of the works
 of the instantiation tasks obtained by the split of the $i$-th literal
with $1\leq i\leq 4$, by varying the target number $s$ of splits. In particular, the first column shows the target
number of splits, the following four columns show the maximum number of splits $s^i$ allowed for each literal,
and the remaining four columns show the costs ${\cal C}^i$ computed according to the $s^i$ values;
in bold face we outline, for each target number of splits, the minimal values of ${\cal C}^i$.
It can be noted that, in our example, increasing the value of $s$
corresponds to different choices of the literal to split.
Moreover, in each row, the choice is always restricted to the first $i$ literals,
where the $i$th literal is the first one allowing for the target number $s$ of splits;
indeed, ${\cal C}^i \leq {\cal C}^k$, for each $k>i$. Furthermore, even a literal that
does not allow for $s$ splits can be chosen; this is the case for $s=100$, where the chosen
literal is $b(Y,Z)$, which allow for $50$ splits.
Notice that the choice of the literal to split may be influenced by the body ordering in some cases,
which in turn considerably affects the serial evaluation time (which is the amount to be divided by parallel evaluation).
For example, all body orderings having $d(V,Z)$ as first literal
have $d$ as chosen literal, since its extension is sufficiently large to allow
the four target numbers of splits considered. However, if such orderings
determine an higher evaluation cost w.r.t. the body order exploited in the serial evaluation,
then the effect of parallel evaluation could be overshadowed.
Thus, we apply the selection of the literal to split after body reordering
with a strategy that minimizes the heuristic cost of instantiating the body.

Summarizing, our heuristics consists in determining an ordering of the body literals
exploiting the already assessed technique described in \cite{leon-etal-2001a}
and splitting the first literal in the body if it allows for the target number of splits (without computing any cost).
Otherwise, the estimations of the costs are determined and the literal allowing for the minimum one is chosen.

\begin{table}
\small
\caption{Number of splits and costs of the instantiation tasks}
\label{funValues}
\begin{tabular}{| c | c | c | c | c | c | c | c | c |}
 \cline{1-9}
$s$ &  $s^1$ &  $s^2$ & $s^3$ &  $s^4$  & ${\cal C}^1$ & ${\cal C}^2$  &${\cal C}^3$ & ${\cal C}^4$ \\
 \cline{1-9}
5 & 5 & 5 & 5 & 5& \bf{11400}  & 11400  & 12200 & 17000 \\
50 & 20 & 50 & 50 & 50 & 2850 & \bf{1140} & 2120 & 8000 \\
100 & 20 & 50 & 100 & 100 & 2850 &  \bf{1140} & 1560 & 7500 \\
500 & 20 & 50 & 500 & 500 & 2850  & 1140  &\bf{1112} & 7100 \\
\cline{1-9}
\end{tabular}

\end{table}

\section{Load Balancing and Granularity Control} \label{sec:loadB}
An advanced parallelization technique has to deal with two important issues that
strongly affect the performance of a real implementation: load balancing and granularity control.
Indeed, if the workload is not uniformly distributed to the available processors
then the benefits of parallelization are not fully obtained; moreover, if the
amount of work assigned to each parallel processing unit is too small then
the (unavoidable) overheads due to creation and scheduling of parallel tasks
might overcome the advantages of parallel evaluation (in corner cases, adopting
a sequential evaluation might be preferable).

As an example, consider the case in which we are running the instantiation of a rule $r$ on a two
processor machine and,
 by applying the technique for Single Rule parallelism described above,
 the instantiation of $r$ is divided into two smaller tasks, by partitioning
 the extension of the split predicate of $r$ into two subsets with,
 approximatively, the same size.
Then, each of the two tasks will be processed by a thread; and the two
threads will possibly run separately on the two available
processors. For limiting the inactivity time of the processors, it
would be desirable that the threads terminate their execution almost
at the same time. Unfortunately, this is not always the
case, because subdividing the extension of the split predicate into
equal parts does not ensure that the workload is equally spread
among threads. However, if we consider a larger number of splits,
a further subdivision of the workload would be obtained, and, the inactivity
time would be more likely limited.

Clearly, it is crucial to guarantee that the parallel evaluation of a number of tasks
is not more time-consuming than their serial evaluation (granularity control); and that
an unbalanced workload distribution does not introduce
significant delays and limits the overall performance (load balancing).

\paragraph{Granularity Control.}
Our method for granularity control is based on the use of a heuristic value \w{r},
which acts as a litmus paper indicating the amount of work required for evaluating each rule of the program,
and so, its ``hardness'', just before its instantiation.
\w{r} denotes the value ${\cal C}(n)$ (see Section \ref{sec:splitlit}), for each rule $r$ having $n$ body literals.

At the rules level, rather than assigning each rule to a different thread, a set of rules $R$ is determined and assigned to a thread.
$R$ is such that the total work for instantiating its rules
is enough for enjoying the benefits of scheduling a new thread.
In practice, $R$ is constructed by iterating on the rules of the same component, and
stopping if either $\sum_{r \in R}\w{r}  > w_{seq}$ or when no further rules can be added to $R$,
where $w_{seq}$ is an empirically-determined threshold value.
At the single rule level, a rule $r$ is scheduled for parallel instantiation (i.e.
its evaluation can be divided into smaller tasks that can be performed in parallel)
if $\w{r}>w_{seq}$; otherwise, for $r$ the third level of parallelism it is not applied at all.

Note that, for simplifying the presentation of the algorithms in Section~\ref{sec:parallel},
we have not considered the management of the granularity control in the second level of parallelism,
which would have added noisy technical details and made the description more involved.
However, they can be suitably adapted by modifying procedure {\em Rules\_Instantiator} (see Fig.\ref{fig:parallelInstantiation1})
in order to build a set of ``easy'' rules, and by adding a {\em SetOfRules\_Instantiator} procedure, which instantiates each rule in the set.
Note also that, granularity control in the third level of parallelism is obtained by setting the number of splits of a given rule to 1.

\paragraph{Load Balancing.} In our approach load balancing exploits different factors.
On the one hand, in the case of the evaluation of recursive rules, a dynamic load redistribution
of the extension of the split literal at each iteration is performed.
On the other hand, the extension of the split literal is divided by a number which
is greater than the number of processors (actually, a multiple of the number of processors is enough)
for exploiting the preemptive multitasking scheduler of the operating system.
Moreover, in case of ``very hard'' rules, a finer distribution is performed in the last splits.
 In particular, when a rule is assessed to be ``hard''
by comparing the estimated work (the value $\w{r}$ described above) with another empirically-determined threshold ($\w{r}>w_{hard}$),
a finer work distribution (exploiting a unary split size) is performed for the last $s-n_{p}$ splits,
where $s$ is the number of splits and $n_{p}$ is the number of processors.
The intuition here is that, if a rule is hard to instantiate then
it is more likely that its splits are also hard, and thus
an uneven distribution of the splits to the available processors
in the last part of the computation might cause a sensible loss of efficiency.
Thus, further subdividing the last ``hard'' splits, may help to distribute in a finer way
the workload in the last part of the computation.

\section{Experiments}\label{sec:experiments}

The parallelization techniques described in the previous sections
were implemented in the instantiation module of DLV~\cite{leon-etal-2002-dlv}.
In order to assess the performance of the resulting parallel instantiator
we carried out an extensive experimental activity, reported in this section.
In particular, $(i)$
we compared the previous techniques (components and rules parallelism) with the new technique
(single-rule parallelism); and $(ii)$ we conducted
a scalability analysis on the instantiator considering  the effects of increasing both the number of available processors
and the size of the instances. Before discussing the results, a description of
the implemented system and some benchmarks data are given.

\subsection{Implementation in \dlv}

We implemented our parallel techniques by extending the instantiator module of \dlv.
The system is implemented in the C{\small ++} language by exploiting the Linux POSIX Thread API,
shipped with the GCC 4.3.3 compiler.
The actual implementation of the algorithms reported in Section~\ref{sec:parallel}
adopts a producer-consumers pattern in which the total number of threads spawned in each
of the three levels of parallelism is limited to a fixed number which is user-defined.
This is obtained by adapting procedures described in Section~\ref{subsec:paroverview},
in such a way that, spawn commands are replaced by push operations in three different shared buffers (one for each level of parallelism);
moreover, thread joins ensuring the completion of a given task (e.g., evaluation of all splits of the same rule)
are replaced by proper condition statements (e.g. counting semaphores).
In this way, worker threads are recycled, and continuously pop working tasks from the buffers, up to the end of the instantiation process.
The main motivation for this technical variant is limiting thread creation overhead to a single initialization step.
In addition, the implementation allows for separately enabling/disabling the three levels of parallelism by command line options.

\subsection{Benchmark Problems and Data}
In our experiments, several well-known combinatorial problems as well as real-world problems are considered.
These benchmarks have been already used for assessing ASP instantiator
performance \cite{leon-etal-2002-dlv,gebs-etal-2007-lpnmr-competition,devebogetr09a}.
Many of them are particularly difficult to parallelize
due to the compactness of their encodings;
note that concise encodings are quite common given
the declarative nature of the ASP language which allows to compactly
encode even very hard problems.
About data, we considered five instances (where the instantiation time is non negligible) of
increasing difficulty for each problem, except for the Hamiltonian Path and
3-Colorability problem, for which generators are available,
and we could generate several instances of increasing size.

In order  to meet space constraints, encodings are not
presented but they are available, together with the employed
instances, and the binaries, at
{\em http://www.mat.unical.it/ricca/ downloads/parallelground10.zip}. Rather,
to help the understanding of the results, both a description of problems and some information
on the number of rules of each program is reported below.

\ \\
\noindent{\bf n-Queens.}
The $8$-queens puzzle is the problem of putting eight chess queens
on an $8$x$8$ chessboard so that none of them is able to capture
any other using the standard chess queen's moves.
The $n$-queens puzzle is the more general problem of
placing $n$ queens on an $n$x$n$ chessboard ($n \geq 4$).
The encoding consists of one rule and four constraints.
Instances were considered having $n \in \{37,39,41,43,45\}$.

\noindent{\bf Ramsey Numbers.}
The Ramsey number\ $ramsey(k,m)$\ is the least integer $n$ such
that, no matter how the edges of the complete undirected graph
(clique) with $n$ nodes are colored using two colors, say red and
blue, there is a red clique with $k$ nodes (a red $k$-clique) or a
blue clique with $m$ nodes (a blue $m$-clique). The encoding of this
problem consists of one rule and two constraints. For the
experiments, the problem was considered of deciding whether, for
$k=7$, $m=7$, and $n\in\{31,32,33,34,35\}$, $n$ is $ramsey(k,m)$.

\noindent{\bf Clique.}
A clique in an undirected graph $G=(V,E)$ is a subset of its vertices such that every two
vertices in the subset are connected by an edge. We considered the problem
finding a clique in a given input graph.
Five graphs of increasing size were considered.

\noindent{\bf Timetabling.\ }
The problem of determining a timetable for some
university lectures which have to be given in a week to some groups
of students. The timetable must respect a number of given
constraints concerning availability of rooms, teachers, and other
issues related to the overall organization of the lectures. Many
instances were provided by the University of Calabria; they refer
to different numbers of student groups $g \in \{15,17,19,21,23\}$.

\noindent{\bf Sudoku.}
Given an $NxN$ grid board, where $N$ is a square number $N=M^2$,
fill it with integers from $1$ to $N$ so that each row, each column,
and each of the $N$ $MxM$ boxes contains each of the integers from $1$ to $N$ exactly once.
Suppose the rows are numbered $1$ to $N$ from left to right, and the columns are
numbered $1$ to $N$ from top to bottom. The boxes are formed by dividing the rows from
top to bottom every $M$ rows and dividing the columns from left to right
every $M$ columns. Encoding and instances were used for testing
the competitors solvers in the ASP Competition 2009~\cite{devebogetr09a}.
For assessing our system we considered the instances
\{$sudoku.in5$, $sudoku.in6$, $sudoku.in7$, $sudoku.in9$, $sudoku.in10$\}, where
$N=25$.

\noindent{\bf Golomb Ruler.}
A Golomb ruler is an assignment of marks to integer positions along a ruler so
that no pair of two marks is the same distance from each other.
The number of marks is the order of the ruler. The first mark is required to be at position 0,
the position of the highest mark is the length of the ruler.
The problem is finding the shortest ruler of a given order.
Encoding and instances have been used for testing
the competitors solvers in the ASP Competition 2009~\cite{devebogetr09a}.
Instances are descibed by a couple $(m,p)$ where  $m$ is the number marks
and $p$ is the number of positions: we considered the values $(10,125)$, $(13,150)$,
$(14,175)$,$(15,200)$ and $(15,225)$.

\noindent{\bf Reachability. } Given a directed graph $G=(V,E)$,
we want to compute all pairs of nodes $(a,b) \in V \times V$ (i)
such that $b$ is reachable from $a$ through a nonempty sequence of
edges in $E$. The encoding of this problem consists of one exit rule
and a recursive one. Five trees were generated
with a pair (number of levels, number of siblings):
(9,3), (7,5), (14,2), (10,3) and (15,2), respectively.

\noindent{\bf Food.\ }
The problem here is to generate plans for repairing faulty
workflows. That is, starting from a faulty workflow instance, the
goal is to provide a completion of the workflow such that the
output of the workflow is correct. Workflows may comprise many
activities. Repair actions are compensation, (re)do and
replacement of activities. A single instance was provided
related to a workflow containing 63 predicates,
56 components and 116 rules.

\noindent{\bf 3-Colorability. } This well-known problem asks for an assignment of
three colors to the nodes of a graph, in such a way that adjacent
nodes always have different colors. The encoding of this problem
consists of one rule and one constraint. A number of simplex graphs were
generated with the Stanford GraphBase library~\cite{knut-94}, by
using the function $simplex(n,n,-2,0,$ $0,0,0)$, where $80\leq n \leq 250$.

\noindent{\bf Hamiltonian Path.}
A classical \NP-complete problem in graph theory, which can be
expressed as follows: given a directed graph $G=(V,E)$ and a node $a
\in V$ of this graph, does there exist a path in $G$ starting at $a$
and passing through each node in $V$ exactly once. The encoding of
this problem consists of several rules, one of these is recursive.
Instances were generated, by  using a tool by Patrik Simons
(cf.~\cite{simo-2000}), with $n$ nodes  with $1000 \leq n \leq 12000$.

The machine used for the experiments is a two-processor Intel Xeon
``Woodcrest'' (quad core) 3GHz machine with 4MB of L2 Cache and 4GB
of RAM,
running Debian GNU Linux 4.0.
Since our techniques focus on instantiation, all the results of the
experimental analysis refer only to the instantiation process
rather than the whole process of computing answer sets;
in addition, the time spent before the
grounding stage (parsing) is obviously the same both for parallel and
non-parallel version. In order to obtain more trustworthy results,
each single experiment was repeated five times and
the average of performance measures are reported.

\begin{figure}[t!]

\subfigure[Average instantiation times in seconds (standard deviation)]{
\scriptsize
\begin{tabular}{| c | c | c | c | c |}
\cline{1-5}
Problem & \serial & \kali & \splitonly & \paral \\
\cline{1-5}

$queen_1$ &  4.64 (0.00) & 2.19 (0.06) & 0.71 (0.01) & 0.69 (0.01) \\
$queen_2$ & 5.65 (0.00) & 3.29 (0.51) & 0.89 (0.01) & 0.86  (0.02) \\
$queen_3$ & 6.83 (0.00)  & 3.85 (0.50) & 1.08 (0.00) & 1.03 (0.02) \\
$queen_4$ & 8.19 (0.00) & 4.50 (0.48) & 1.25 (0.01) & 1.22 (0.00) \\
$queen_5$ & 9.96 (0.00) & 4.72 (0.11) & 1.45 (0.02) & 1.43 (0.01) \\
\cline{1-5}
$ramsey_1$ & 258.52 (0.00) & 131.61 (0.23) & 40.53 (0.21) & 36.23 (0.34)\\
$ramsey_2$ & 328.68 (0.00) & 168.03 (1.38) & 51.35 (0.25) & 46.09 (0.33)\\
$ramsey_3$ & 414.88 (0.00) & 211.60 (0.55) & 68.49 (0.10) & 58.06 (0.20)\\
$ramsey_4$ & 518.28 (0.00) & 265.58 (2.82) & 83.32 (0.24) & 75.19 (2.41)\\
$ramsey_5$ & 643.65 (0.00) & 327.25 (0.79) & 103.14 (0.44) & 92.28 (0.65)\\
\cline{1-5}
$clique_1$ & 16.06 (0.00) & 15.88 (0.20) & 3.34 (0.04) & 2.35 (0.01)\\
$clique_2$ & 29.98 (0.00) & 29.97 (0.01) & 4.41 (0.12) & 4.34 (0.07)\\
$clique_3$ & 49.11 (0.00) & 49.18 (0.05) & 7.11 (0.03) & 7.09 (0.02) \\
$clique_4$ & 78.05 (0.00) & 78.70 (0.35) & 11.35 (0.14) & 11.29 (0.11) \\
$clique_5$ & 119.48 (0.00) & 118.66 (0.07) & 17.08 (0.14) & 17.09 (0.16) \\
\cline{1-5}
$timetab_1$ & 15.48 (0.00) & 7.28 (0.04) & 2.35 (0.04) & 2.29 (0.01)\\
$timetab_2$ & 17.49 (0.00) & 8.30 (0.07) & 2.61 (0.01) & 2.61 (0.02)\\
$timetab_3$ & 21.65 (0.00) & 10.22 (0.05) & 3.22 (0.04) & 3.20 (0.01)\\
$timetab_4$ & 17.75 (0.00) & 8.24 (0.04) & 2.61 (0.01) & 2.64 (0.05)\\
$timetab_5$ & 23.69 (0.00) & 11.09 (0.01) & 3.52 (0.01) & 3.50 (0.03)\\
\cline{1-5}
$sudoku_1$ & 5.42 (0.00) & 4.15 (0.04) & 0.98 (0.01) & 0.88 (0.01)\\
$sudoku_2$ & 9.87 (0.00) & 7.75 (0.01) & 1.59 (0.02) & 1.51 (0.02)\\
$sudoku_3$ & 10.28 (0.00) & 8.01 (0.05) & 1.62 (0.00) & 1.57 (0.01)\\
$sudoku_4$ & 10.56 (0.00) & 8.38 (0.01) & 1.75 (0.02) & 1.63 (0.03)\\
$sudoku_5$ & 11.08 (0.00) & 8.25 (0.04) & 1.67 (0.02) & 1.63 (0.05)\\
\cline{1-5}
$gol\_ruler_1$ & 6.58 (0.00) & 6.32 (0.07) & 0.96 (0.01) & 0.94 (0.02)\\
$gol\_ruler_2$ & 13.74 (0.00) & 12.57 (0.04) & 1.87 (0.04) & 1.84 (0.09)\\
$gol\_ruler_3$ & 24.13 (0.00) & 22.67 (0.05) & 3.29 (0.06) & 3.25 (0.13)\\
$gol\_ruler_4$ & 40.64 (0.00) & 37.50 (0.10) & 5.44 (0.21) & 5.51 (0.10)\\
$gol\_ruler_5$ & 62.23 (0.00) & 59.04 (0.04) & 8.32 (0.12) & 8.36 (0.17)\\
\cline{1-5}
$reach_1$ & 52.21 (0.00) & 52.07 (0.07) & 8.25 (0.06) & 8.28 (0.01) \\
$reach_2$ & 147.34 (0.00) & 148.34 (0.01) & 22.60 (0.16) & 22.67 (0.18) \\
$reach_3$ & 258.01 (0.00) & 240.17 (0.13) & 39.59 (0.29) & 39.57 (0.44) \\
$reach_4$ & 522.09 (0.00) & 517.97 (0.59) & 77.21 (0.20) & 77.52 (0.31) \\
$reach_5$ & 1072.00 (0.00) & 1069.86 (1.04) & 160.66 (0.21) & 160.31 (0.25) \\
\cline{1-5}
$Food$& 684.95 (1.19)  & 0.18 (0.15) & 104.6 (1.01) & 0.08 (0.01) \\
\cline{1-5}
\end{tabular}
 \label{tab:tec}
}
\centering
\subfigure[Instantiation times(s) - Hamiltonian Path ]{\includegraphics[scale=0.23,angle=270]{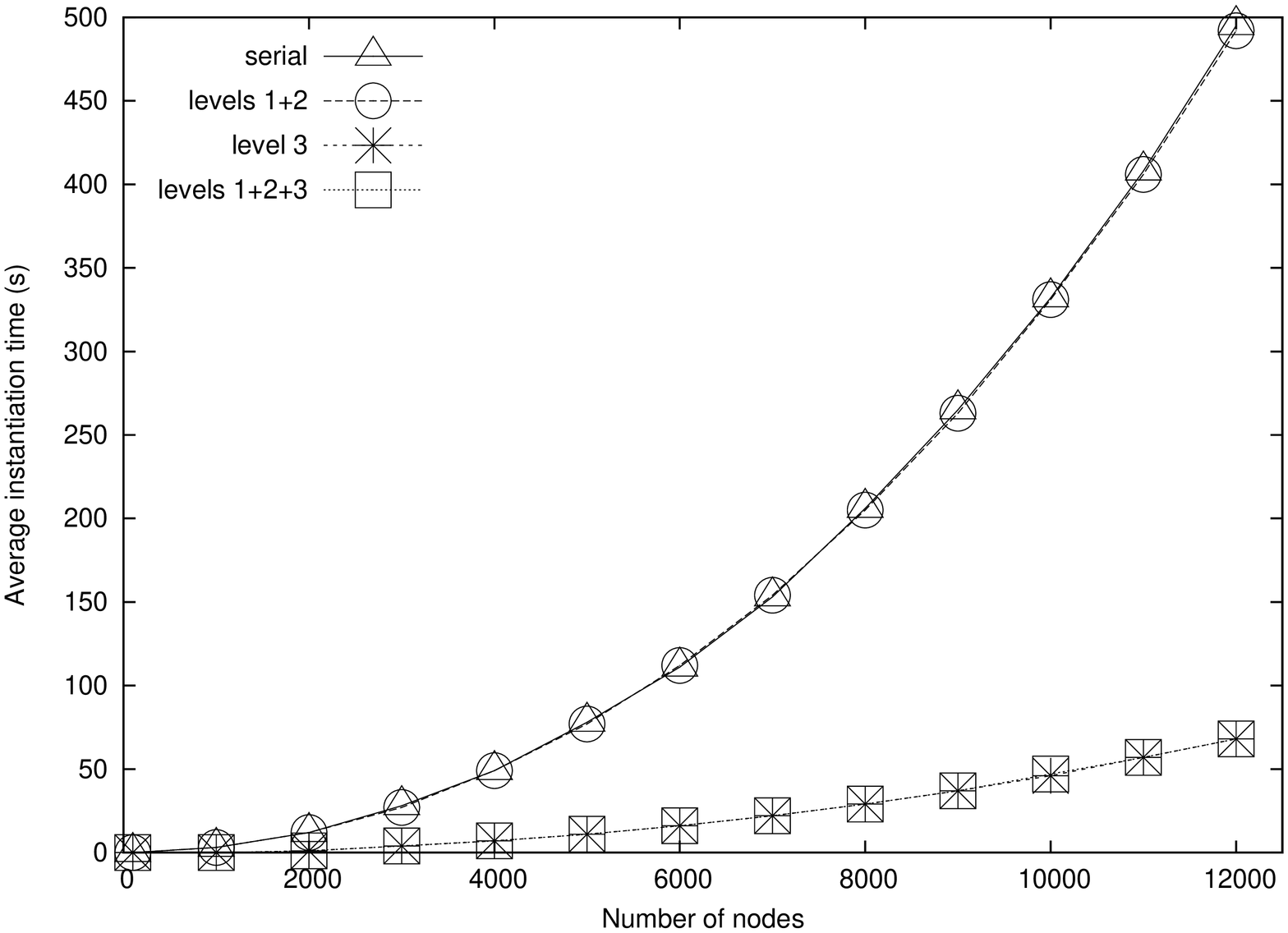}\label{fig:hptec}}
\hspace{0.0cm}
\subfigure[Instantiation times(s) - 3-Colorability ]{\includegraphics[scale=0.22,angle=270]{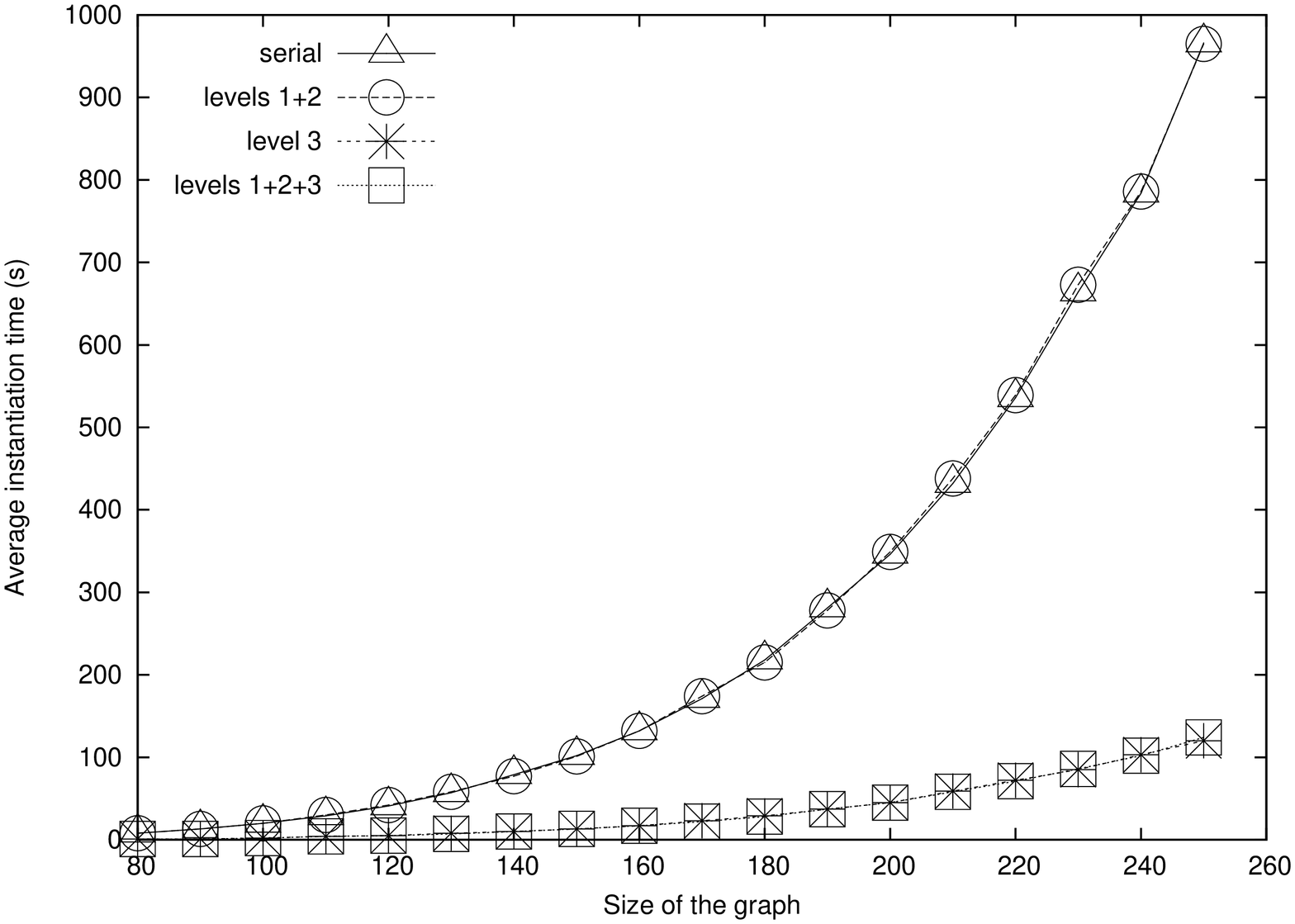}\label{fig:3coltec}}
\hspace{5mm}
\caption{Impact of parallelization techniques} \label{tab:technique}
\end{figure}

\subsection{Effect of Single Rule Parallelism}

In this section we report the results of an experimental analysis aimed at comparing the effects
of the single rule parallelism with the first two levels.
More in detail, we considered four versions of the instantiator: ($i$) \serial where parallel techniques are not applied,
$(ii)$ \kali where components and rules parallelism are applied,
$(iii)$ \splitonly where only the single rule level is applied, and
$(iv)$ \paral in which all the three levels are applied.
Results are shown in Fig.~\ref{tab:technique}(a)(b)(c), where the two graphs report
the average instantiation times for the Hamiltonian Path problem and 3-Colorability; while the table reports
the average instantiation times in seconds for the remaining benchmarks.%
\footnote{Results in form of tables for Hamiltonian Path and 3-Colorability are reported in Appendix B.}

More in detail, in Fig.~\ref{tab:tec},
the first column reports the problem considered, whereas the next columns report
the results for the four instantiators.
Looking at the third column in the table, benchmarks can be classified in three different groups
according to their behaviors: the benchmarks in which the first two levels of parallelism
apply, those where these first two levels apply marginally, and those where they do not apply at all.
In the first group, we have the n-Queens problem, Ramsey Numbers, and Timetabling,
where \kali is about twice faster than $\serial$; however,
considering that the machine on which we ran the benchmarks has eight core available,
$\kali$ is not able to exploit all the computational power at hand.
The reason, is that the encodings of these benchmarks either have a small number of rules
(n-Queens, Ramsey Numbers), or they show an intrinsic dependency among
components/rules (Timetabling), that limits the efficacy of the first two levels of parallelism.
These considerations explain also the behavior of the other two groups of benchmarks. More in detail, for the second group (which contains only Sudoku)
a small improvement is obtained due to few rules which are evaluated in parallel, while the benchmarks belonging to the third group, whose encodings have very few interdependent rules (e.g. Reachability), proved hard to parallelize. Looking at the graphs, Hamiltonian Path and 3-Colorability clearly belongs to the third group, indeed the lines of \serial and \kali overlap.

A special case is the Food problem, showing an impressive performance,
which proved to be a case very easy to parallelize.
This behavior can be explained by a different scheduling of the constraints performed
by the serial version and the \kali one. In particular, this instance is inconsistent
(there is a constraint always violated) and both versions stop the
computation as soon as they recognize this fact. The scheduling
performed by the parallel version allows the identification of this situation before
the serial one, since constraints are evaluated in parallel, while the latter
evaluates the inconsistent constraint later on.

Concerning the behavior of \splitonly, we notice that it always performs very well (always more than 7.5x faster than \serial), and
in all cases but Food, it outperforms \kali.
Basically, the third level of parallelism applies to every single rule, and thus it is effective on all problems, even those with very small encodings.
In the case of Food, even if \splitonly is about 7x faster than \serial, it evaluates rules in the same order than \serial thus recognizing the inconsistent constraint
later than \kali.

The good news is that the three levels of parallelism always combine
(even in the case of Food). This can be easily seen by looking at the last column of table and at the two graphs.
Note that most of the advantages are due to the third level of parallelism. Indeed in the graphs, the lines
for \splitonly and \paral overlap, and \paral shows only marginal gains w.r.t. \splitonly, in the benchmarks
where \kali applies.

\subsection{Scalability of the Approach}
\begin{figure*}[t]
\centering
\subfigure[Efficiency - Hamiltonian Path  ]{\includegraphics[scale=0.23,angle=270]{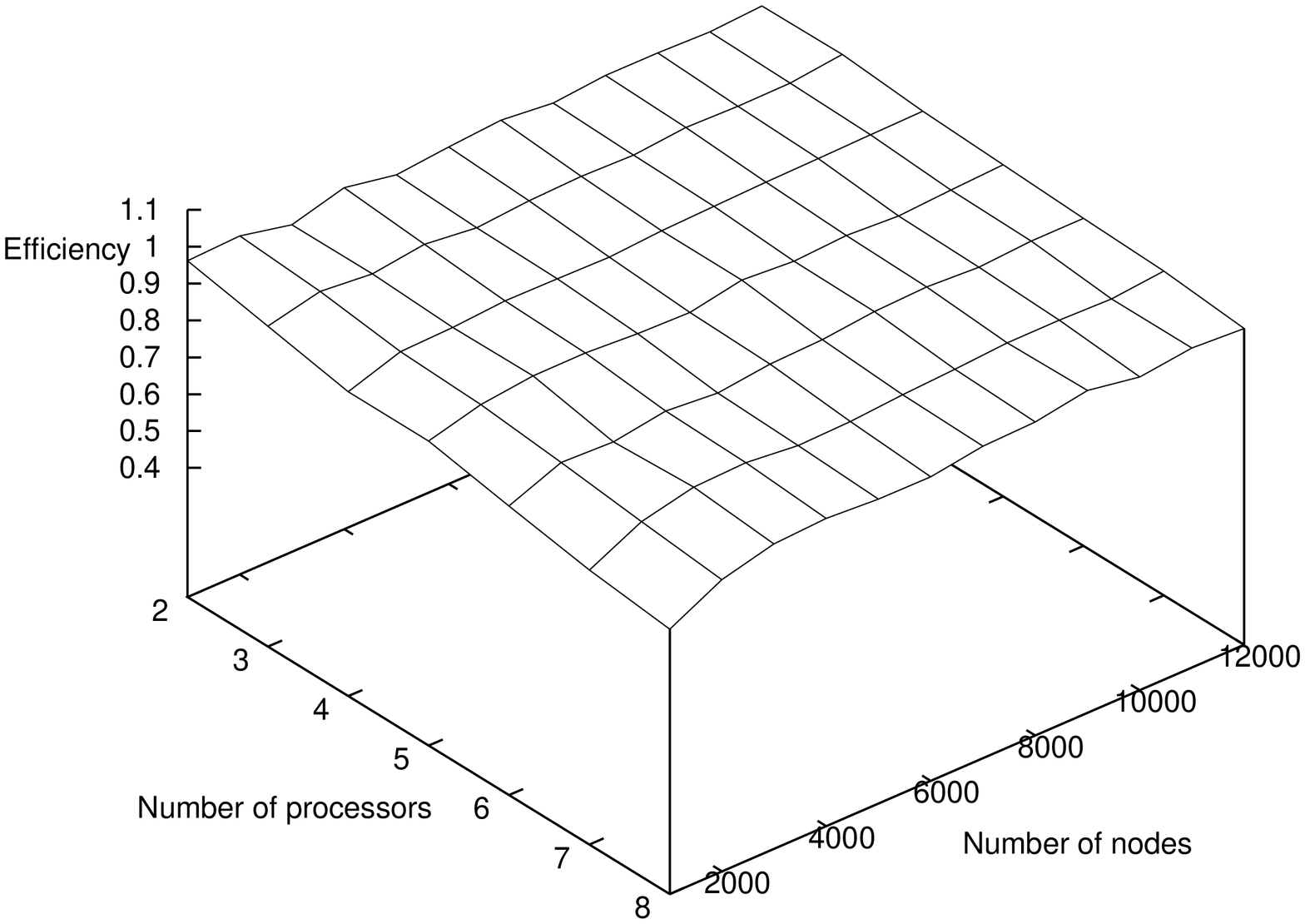}\label{fig:hpEff}}
\hspace{0.0cm}
\subfigure[Instantiation times - Hamiltonian Path ]{\includegraphics[scale=0.22,angle=270]{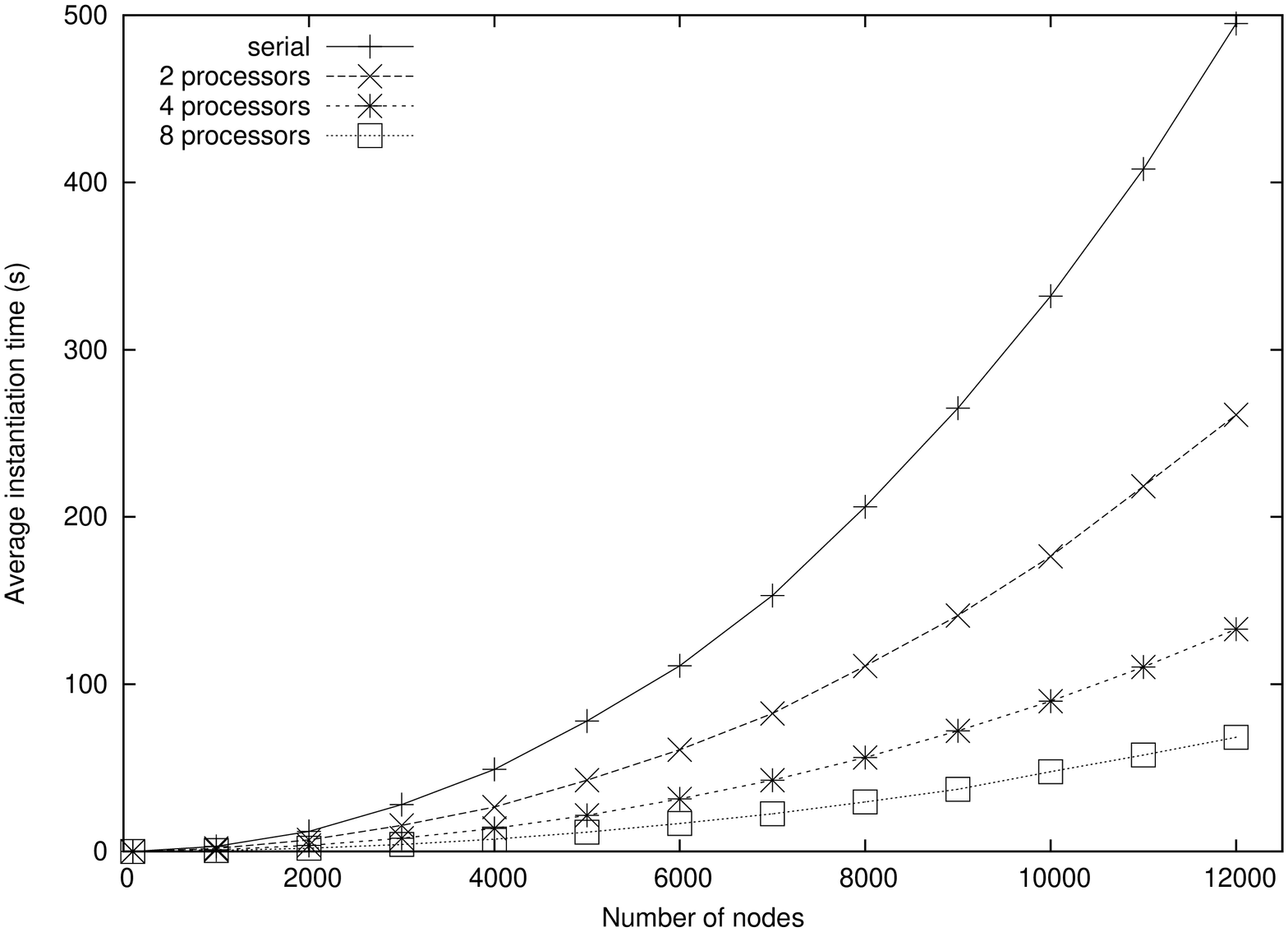}\label{fig:hpTime}}
\hspace{5mm}
\centering
\subfigure[Efficiency - 3-Colorability ]{\includegraphics[scale=0.23,angle=270]{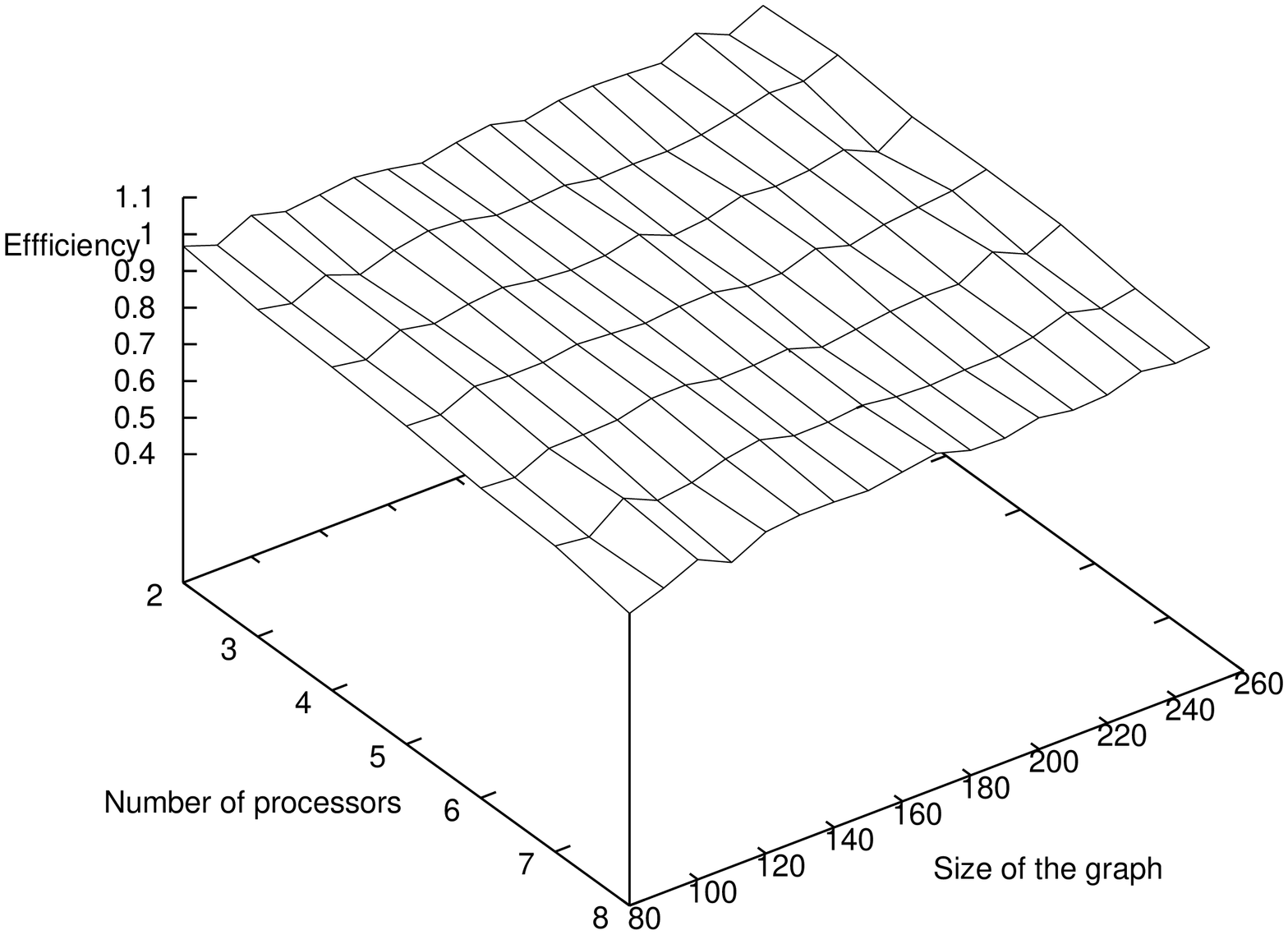}\label{fig:3colEff}}
\hspace{0.0cm}
\subfigure[ Instantiation times - 3-Colorability ]{\includegraphics[scale=0.22,angle=270]{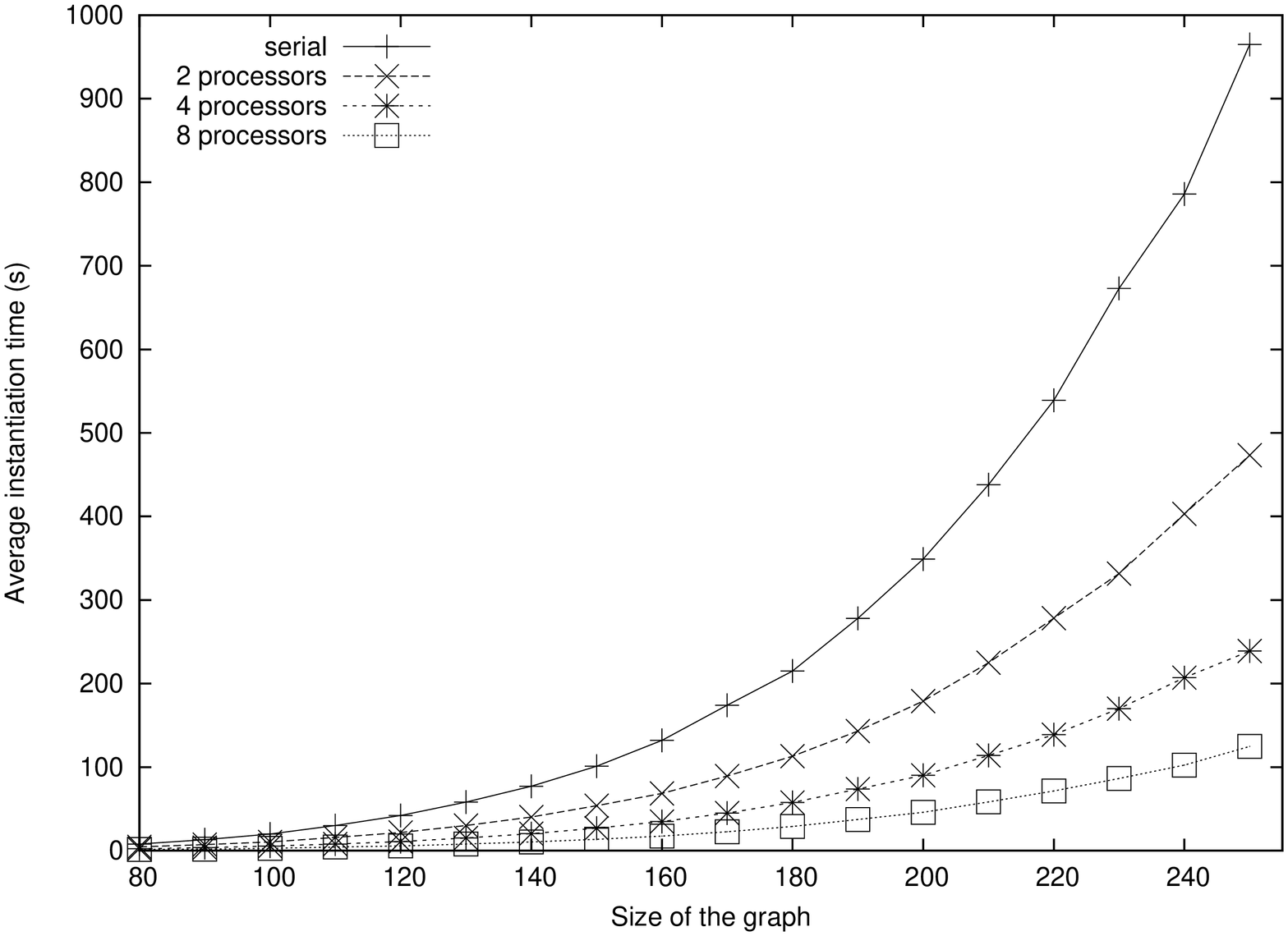}\label{fig:3colTime}}
\hspace{5mm}
\caption{Hamiltonian Path and 3-Colorability: average instantiation times (s) and efficiency}\label{fig:scalability}
\end{figure*}

We conducted a scalability analysis on the instantiator \paral which exploits all the three parallelism levels.
Moreover, we considered the effects of increasing both the size of the instances and the number of available processors (from 1 up to 8 CPUs).%
\footnote{
Available processors can be disabled (respectively enabled) by running the bash Linux commands:
\mbox{{\em echo 0 $>>$ /sys/devices/system/cpu/cpu-n/online}} (resp. \mbox{{\em echo 1 $>>$ /sys/devices/system/cpu/cpu-n/online}}).
}
The results of  the analysis are summarized in Table~\ref{table2}, and
Figure~\ref{fig:scalability}, where both the average instantiation times and the efficiencies are reported.
As before, the graphs show results for Hamiltonian Path and 3-Colorability,
while the results of the remaining problems are reported in the table.
The overall picture is very positive: the performance of the instantiator
is very good in all cases and average efficiencies vary in between 0.85 and 0.95 when all the available CPUs are enabled.
As one would expect, the efficiency of the system both slightly decreases when the number of processors
increases --still remaining at a good level-- and rapidly increases going from very small instances
($<$2 seconds) to larger ones.

The granularity control mechanism resulted to be effective in the n-Queens problem,
where all the considered instances required less than 10 seconds of serial execution time.
Indeed, the ``very easy'' disjunctive rule was always sequentially-evaluated in all the cases.
Since the remaining constraints are strictly determined by the result of the evaluation
of the disjunctive rule, the unavoidable presence of a sequential part limited
the final efficiency to a remarkable 0.9 in the case of 8 processors.
A similar scenario can be observed in the case of Ramsey Numbers,
where the positive impact of the load balancing and granularity
control heuristics becomes very evident.
In fact, since the encoding is composed of few ``very easy'' disjunctive rules
and two ``very hard'' constraints, the heuristics selects a sequential evaluation
for the rules, and dynamically applies the finer distribution of the last splits for
%
\begin{landscape}

\scriptsize
\begin{tabular}{@{\extracolsep{\fill}}|c|r|r|r|r|r|r|r|r||r|r|r|r|r|r|r|}
\cline{1-16}
&\multicolumn{8}{c||}{Average instantiation time (standard deviation)}& \multicolumn{7}{c|}{Efficiency}\\
\cline{1-16}
Problem & serial & 2 proc  & 3 proc & 4 proc & 5 proc & 6 proc & 7 proc & 8 proc &  2 proc  & 3 proc & 4 proc & 5 proc & 6 proc & 7 proc & 8 proc\\
\cline{1-16}
$queen_1$ & 4.64 (0.00) & 2.53 (0.01) & 1.71 (0.01) & 1.31 (0.01) & 1.07 (0.00) & 0.91 (0.01) & 0.78 (0.01) & 0.69 (0.01)&0.98 & 0.97 & 0.95 & 0.93 & 0.91 & 0.91 & 0.90\\
$queen_2$ & 5.65 (0.00) & 3.11 (0.00) & 2.11 (0.01) & 1.60 (0.00) & 1.31 (0.01) & 1.11 (0.01) & 0.97 (0.00) & 0.86 (0.02)&0.99 & 0.97 & 0.96 & 0.94 & 0.92 & 0.91 & 0.90\\
$queen_3$ & 6.83 (0.00) & 3.79 (0.01) & 2.57 (0.01) & 1.97 (0.01) & 1.60 (0.01) & 1.35 (0.02) & 1.17 (0.01) & 1.03 (0.02)&0.99 & 0.97 & 0.95 & 0.94 & 0.92 & 0.91 & 0.91\\
$queen_4$ & 8.19 (0.00) & 4.54 (0.00) & 3.06 (0.01) & 2.35 (0.01) & 1.90 (0.01) & 1.62 (0.02) & 1.41 (0.01) & 1.22 (0.00)&0.99 & 0.98 & 0.96 & 0.95 & 0.92 & 0.91 & 0.92 \\
$queen_5$ &  9.96 (0.00) & 5.57 (0.19) & 3.68 (0.01) & 2.81 (0.02) & 2.26 (0.02) & 1.92 (0.00) & 1.69 (0.01) & 1.43 (0.01)&0.97 & 0.97 & 0.96 & 0.95 & 0.93 & 0.94 & 0.94 \\
\cline{1-16}
$ramsey_1$ & 258.52 (0.00) & 131.72 (0.08) & 89.04 (0.41) & 67.10 (0.46) & 55.14 (0.93) & 46.62 (0.19) & 39.98 (0.12) & 36.23 (0.34)& 0.98 & 0.97 & 0.96 & 0.94 & 0.92 & 0.92 & 0.89 \\
$ramsey_2$ & 328.68 (0.00) & 167.47 (0.16) & 112.97 (0.94) & 85.90 (0.15) & 70.64 (1.74) & 58.70 (0.82) & 51.21 (0.18) & 46.09 (0.33)& 0.98 & 0.97 & 0.96 & 0.93 & 0.93 & 0.92 & 0.89 \\
$ramsey_3$ & 414.88 (0.00) & 210.98 (0.38) & 142.85 (0.68) & 108.00 (0.38) & 88.13 (0.51) & 74.83 (0.22) & 65.25 (0.59) & 58.06 (0.20)& 0.98 & 0.97 & 0.96 & 0.94 & 0.92 & 0.91 & 0.89 \\
$ramsey_4$ & 518.28 (0.00) & 264.69 (1.82) & 178.67 (2.39) & 137.42 (1.89) & 111.09 (2.15) & 95.27 (2.02) & 81.45 (0.45) & 75.19 (2.41)& 0.98 & 0.97 & 0.94 & 0.93 & 0.91 & 0.91 & 0.86\\
$ramsey_5$ & 643.65 (0.00) & 327.06 (0.36) & 222.81 (0.20) & 169.37 (0.86) & 135.94 (0.17) & 115.78 (0.92) & 101.21 (1.33) & 92.28 (0.65)& 0.98 & 0.96 & 0.95 & 0.95 & 0.93 & 0.91 & 0.87 \\
\cline{1-16}
$clique_1$ & 16.06 (0.0) & 8.51 (0.13) & 5.84 (0.08) & 4.45 (0.17) & 3.64 (0.04) & 3.08(0.1) & 2.67 (0.03) & 2.35 (0.01)&0.94 & 0.92 & 0.90 & 0.88 & 0.87 & 0.86 & 0.85 \\
$clique_2$ & 29.98 (0.0) & 15.92 (0.23) & 10.69 (0.18) & 8.27 (0.09) & 6.77 (0.11) & 5.71 (0.10) & 4.94 (0.40) & 4.34 (0.07)&0.94 & 0.93 & 0.91 & 0.89 & 0.88 & 0.87 & 0.86 \\
$clique_3$ & 49.11 (0.00) & 25.81 (0.41) & 17.31 (0.06) & 13.39 (0.20) & 10.92 (0.20) & 9.23 (0.02) & 7.98 (0.03) & 7.09 (0.02)& 0.95 & 0.95 & 0.92 & 0.90 & 0.89 & 0.88 & 0.87 \\
$clique_4$ & 78.05 (0.00) & 41.68 (0.07) & 27.91 (0.28) & 21.10 (0.02) & 17.33 (0.20) & 14.60 (0.04) & 12.76 (0.06) & 11.29 (0.11)& 0.94 & 0.93 & 0.92 & 0.90 & 0.89 & 0.87 & 0.86 \\
$clique_5$ & 119.48 (0.00) & 62.87 (0.13) & 42.62 (0.15) & 32.46 (0.04) & 26.14 (0.21) & 22.24 (0.00) & 19.14 (0.00) & 17.09 (0.16)& 0.95 & 0.93 & 0.92 & 0.91 & 0.90 & 0.89 & 0.87 \\
\cline{1-16}
$timetab_1$ & 15.48 (0.00) & 7.98 (0.10) & 5.41 (0.02) & 4.16 (0.00) & 3.37 (0.00) & 2.93 (0.05) & 2.59 (0.03) & 2.29 (0.01)& 0.97 & 0.95 & 0.93 & 0.92 & 0.88 & 0.85 & 0.84  \\
$timetab_2$ & 17.49 (0.00) & 9.26 (0.34) & 6.30 (0.23) & 4.68 (0.01) & 3.89 (0.04) & 3.41 (0.14) & 2.92 (0.04) & 2.61 (0.02)& 0.94 & 0.93 & 0.93 & 0.90 & 0.85 & 0.86 & 0.84 \\
$timetab_3$ & 21.65 (0.00) & 11.12 (0.03) & 7.54 (0.02) & 5.98 (0.23) & 4.84 (0.09) & 4.08 (0.05) & 3.64 (0.02) & 3.20 (0.01)& 0.97 & 0.96 & 0.91 & 0.89 & 0.88 & 0.85 & 0.85 \\
$timetab_4$ & 17.75 (0.00) & 9.32 (0.36) & 6.13 (0.02) & 4.75 (0.06) & 3.86 (0.02) & 3.33 (0.01) & 3.01 (0.15) & 2.64 (0.05)& 0.95 & 0.97 & 0.93 & 0.92 & 0.89 & 0.84 & 0.84 \\
$timetab_5$ & 23.69 (0.00) & 12.16 (0.01) & 8.28 (0.01) & 6.35 (0.02) & 5.34 (0.20) & 4.47 (0.03) & 3.94 (0.03) & 3.50 (0.03)& 0.97 & 0.95 & 0.93 & 0.89 & 0.88 & 0.86 & 0.85 \\
\cline{1-16}
$sudoku_1$ & 5.42 (0.00) & 2.84 (0.01) & 2.14 (0.21) & 1.54 (0.00) & 1.29 (0.02) & 1.10 (0.00) & 0.98 (0.02) & 0.88 (0.01)& 0.95 & 0.84 & 0.88 & 0.84 & 0.82 & 0.79 & 0.77 \\
$sudoku_2$ & 9.87 (0.00) & 5.09 (0.02) & 3.53 (0.02) & 2.72 (0.04) & 2.25 (0.01) & 1.90 (0.02) & 1.68 (0.03) & 1.51 (0.02)& 0.94 & 0.94 & 0.91 & 0.89 & 0.87 & 0.83 & 0.82 \\
$sudoku_3$ & 10.28 (0.00) & 5.45 (0.17) & 3.63 (0.01) & 2.81 (0.02) & 2.31 (0.01) & 1.96 (0.01) & 1.78 (0.00) & 1.57 (0.01)& 0.97 & 0.93 & 0.91 & 0.88 & 0.87 & 0.84 & 0.82 \\
$sudoku_4$ & 10.56 (0.00) & 5.50 (0.03) & 3.80 (0.03) & 2.92 (0.03) & 2.41 (0.02) & 2.03 (0.02) & 1.81 (0.04) & 1.63 (0.03)& 0.96 & 0.93 & 0.90 & 0.88 & 0.87 & 0.83 & 0.81 \\
$sudoku_5$ & 11.08 (0.00) & 5.52 (0.12) & 3.73 (0.01) & 2.93 (0.11) & 2.35 (0.01) & 2.05 (0.04) & 1.82 (0.03) & 1.63 (0.05)& 1.00 & 0.99 & 0.95 & 0.94 & 0.90 & 0.87 & 0.85 \\
\cline{1-16}
$gol\_ruler_1$ & 6.58 (0.00) & 3.34 (0.01) & 2.26 (0.00) & 1.73 (0.02) & 1.42 (0.02) & 1.24 (0.02) & 1.06 (0.03) & 0.94 (0.02)& 0.99 & 0.97 & 0.95 & 0.93 & 0.88 & 0.89 & 0.88 \\
$gol\_ruler_2$ & 13.74 (0.00) & 6.63 (0.02) & 4.60 (0.18) & 3.41 (0.04) & 2.86 (0.10) & 2.43 (0.04) & 2.11 (0.02) & 1.84 (0.09)& 1.04 & 1.00 & 1.01 & 0.96 & 0.94 & 0.93 & 0.93\\
$gol\_ruler_3$ & 24.13 (0.00) & 12.11 (0.02) & 8.15 (0.06) & 6.34 (0.06) & 5.06 (0.10) & 4.34 (0.17) & 3.79 (0.05) & 3.25 (0.13)& 1.00 & 0.99 & 0.95 & 0.95 & 0.93 & 0.91 & 0.93\\
$gol\_ruler_4$ & 40.64 (0.00) & 20.27 (0.05) & 13.51 (0.11) & 10.35 (0.10) & 8.64 (0.19) & 7.13 (0.25) & 6.35 (0.31) & 5.51 (0.10)& 1.00 & 1.00 & 0.98 & 0.94 & 0.95 & 0.91 & 0.92 \\
$gol\_ruler_4$ & 62.23 (0.00) & 31.54 (0.29) & 21.30 (0.16) & 16.03 (0.09) & 12.95 (0.20) & 11.03 (0.27) & 9.67 (0.15) & 8.36 (0.17)& 0.99 & 0.97 & 0.97 & 0.96 & 0.94 & 0.92 & 0.93 \\
\cline{1-16}
$reach_1$ & 52.21 (0.00) & 29.52 (0.36) & 20.41 (0.25) & 15.38 (0.02) & 12.73 (0.16) & 10.81 (0.03) & 9.63 (0.05) & 8.28 (0.01)& 0.86 & 0.85 & 0.85 & 0.82 & 0.80 & 0.77 & 0.79 \\
$reach_2$ & 147.34 (0.00) & 84.93 (0.35) & 57.14 (0.07) & 43.56 (0.07) & 35.16 (0.19) & 29.90 (0.02) & 26.02 (0.10) & 22.67 (0.18)& 0.87 & 0.86 & 0.85 & 0.84 & 0.82 & 0.81 & 0.81 \\
$reach_3$ & 258.01 (0.00) & 144.36 (0.60) & 97.06 (0.39) & 74.88 (0.25) & 61.05 (0.21) & 52.38 (0.09) & 45.64 (0.21) & 39.57 (0.44)& 0.87 & 0.87 & 0.86 & 0.85 & 0.82 & 0.81 & 0.82 \\
$reach_4$ & 522.09 (0.00) & 301.44 (0.48) & 201.90 (0.47) & 153.00 (0.28) & 123.83 (0.16) & 104.74 (0.54) & 90.44 (0.10) & 77.52 (0.31))& 0.87 & 0.86 & 0.85 & 0.84 & 0.83 & 0.82 & 0.84 \\
$reach_5$ & 1072.00 (0.00) & 618.68 (0.71) & 412.29 (0.05) & 311.96 (0.98) & 253.22 (0.98) & 213.31 (0.85) & 185.08 (0.26) & 160.31 (0.25)& 0.87 & 0.87 & 0.86 & 0.85 & 0.84 & 0.83 & 0.84 \\
\cline{1-16}
$Food$& 684.95 (1.19)  & 0.22 (0.15) &  0.08 (0.01) & 0.07 (0.01) & 0.06 (0.01) &  0.06 (0.00) & 0.06 (0.00) & 0.08 (0.01) & 1556 & 2853 & 2446 & 2283 & 1902 & 1630 & 1223 \\
 \cline{1-16}
\end{tabular}
 \captionof{table}{Benchmark Results: average instantiation times in seconds (standard deviation), efficiency}
  \label{table2}
\end{landscape}
%
\noindent the constraints. As a result, the system produces a well-balanced work
subdivision, that allows for obtaining steady results with an average efficiency
greater of almost equal to 0.9 in all tested configurations.
Analogously for Clique, which has a short encoding consisting of only three easy rules, for which
granularity control schedules a serial execution, and one ``hard''
constraint which can be split and thus evaluated in parallel.

A good performance is also
obtained in the case of Reachability. This problem is made up of only
two rules; the first one is caught by granularity control
which schedules its serial execution. The second one is a heavy recursive rule,
that requires several iterations to be grounded.
In this case a good load
balancing is obtained thanks to the
redistributions applied (with possibly different split sizes)
at each iteration of the semi na\"ive algorithm.

The instantiator is effective
also in Golomb Ruler, Timetabling and Sudoku where
the performance results to be good also
thanks to a well-balanced workload distribution.

About Food, a super-linear speedup (due to the first levels of parallelism) is already evident with two-processors and
efficiency peaks when three processors are enabled, where the execution times becomes negligible.
The behavior of the system for instances of varying sizes
was analyzed in more detail in the case of Hamiltonian Path and 3-Colorability;
this was made possible by the availability of generators.

Looking at Figures~\ref{fig:hpEff} and~\ref{fig:3colEff}, it is evident that the efficiency of the system
rapidly reaches a good level (ranging from  0.9 up to 1), moving from small instances
(requiring less than 2s) to larger ones, and remains stable
(the surfaces are basically plateaux).
The corresponding gains are visible by looking at Figures~\ref{fig:hpTime} and ~\ref{fig:3colTime},
where, e.g. an Hamiltonian Path (3-Colorability) instance is evaluated in 332.78s (965.36s) by the serial system,
and requires only 68.26s (124.70s) with \paral with 8-processor enabled.

Summarizing, the parallel instantiator behaved very well in all the considered
instances. It showed superlinear speedups in the case of easy-to-parallelize instances
and, in the other cases its efficiency rapidly reaches good levels and remains stable
when the sizes of the input problem grow. Importantly, the system offers a very good
performance already when only two CPUs are enabled (i.e. for the largest majority of the
commercially-available hardware at the time of this writing), and efficiency remains
at a very good level when up to 8 CPUs are available.

\section{Related Work}\label{sec:related}

Several works about parallel techniques for the evaluation of ASP
programs have been proposed, focusing on both
the propositional (model search) phase%
~\cite{fink-etal-2001,ellg-etal-2009,gres-etal-2005,pont-elkh-2001},
and  the instantiation
phase~\cite{bald-etal-2005-parallel,cali-etal-2008-joacil}. Model
generation is a distinct phase of  ASP computation, carried out
after the instantiation, and thus, the first group of proposals is
not directly related to our setting.
Concerning the parallelization of the instantiation phase, some preliminary
studies were carried out in~\cite{bald-etal-2005-parallel}, as one of the
aspects of the attempt to introduce parallelism in non-monotonic
reasoning systems. However, there are crucial differences with our
system regarding both the employed technology and the supported
parallelization strategy. Indeed, our system  is implemented by
using POSIX threads APIs, and works in a shared memory
architecture~\cite{stall-98}, while the one described
in~\cite{bald-etal-2005-parallel} is actually a
Beowulf cluster working in local memory. Moreover,
the parallel instantiation strategy
of~\cite{bald-etal-2005-parallel} is applicable only to a subset of
the program rules (those not defining domain predicates), and is, in
general, unable to exploit parallelism fruitfully in the case of
programs with a small number of rules. Importantly, the
parallelization strategy of ~\cite{bald-etal-2005-parallel} {\em
statically} assigns a rule per processing unit; whereas, in our
approach, both the extension of predicates and split sizes are
dynamically computed (and updated at different iterations of the
semi-na\"ive evaluation) while the instantiation process is running. Note also
that our parallelization techniques could be
adapted for improving other ASP instantiators
like Lparse~\cite{niem-simo-97} and Gringo~\cite{gebs-etal-2007-gringo}.
Concerning other related works, it is worth remembering that, the
Single Rule parallelism employed in our system is related to the
{\em copy and constrain} technique for parallelizing the evaluation of deductive
databases~\cite{wolf-silb-1988,wolf-ozer-90,gang-etal-90,zhan-etal-1995,dewa-etal-94}.
In many of the mentioned works (dating back to 90's), only restricted classes of
Datalog programs are parallelized; whereas, the most general ones
(reported in \cite{wolf-ozer-90,zhan-etal-1995}) are applicable to
normal Datalog programs. Clearly, none of them consider the
peculiarities of disjunctive programs and unstratified negation.
More in detail, \cite{wolf-ozer-90} provides the theoretical
foundations for the {\em copy and constrain} technique,
whereas~\cite{zhan-etal-1995} enhances it in such a way that
the network communication overhead in distributed systems can be minimized.
The copy and constrain technique works as follows: rules are replicated
with additional constraints attached to each copy;
such constraints are generated by exploiting a hash function
and allow for selecting a subset of the tuples.
The obtained restricted rules are evaluated in parallel. The technique employed in our system
shares the idea of splitting the instantiation of each
rule, but has several differences that allow for obtaining an
effective implementation. Indeed,
in~\cite{wolf-ozer-90,zhan-etal-1995} copied rules are generated and
statically associated to instantiators according to an hash function
which is independent of the current instance in input. In contrast,
in our technique, the distribution of predicate extensions is
performed dynamically, before assigning  the rules to instantiators,
by taking into account the ``actual'' predicate extensions. In this
way, the non-trivial problem~\cite{zhan-etal-1995} of choosing an
hash function that properly distributes the load is completely
avoided in our approach. Moreover, the evaluation of conditions
attached to the rule bodies during the instantiation phase
would require to modify either the standard instantiation procedure
(for efficiently selecting the tuples from the predicate extensions according
to added constraints) or to incur a possible non negligible overhead
due to their evaluation.
Focusing on the {\em heuristics} employed on parallel databases, we
mention~\cite{dewa-etal-94} and \cite{carey-et-all-86}.
In~\cite{dewa-etal-94}  a heuristics is described for balancing the
distribution of load in the parallel evaluation of PARULEL, a
language similar to Datalog. Here, load balancing is done by a
manager server that records the execution times at each site, and
exploits this information for distributing the load according to
predictive dynamic load balancing (PDLB) protocols
that ``update and reorganize the distribution of workload at runtime
by modifying the restrictions on versions of the rule program''\cite{dewa-etal-94}.
In~\cite{carey-et-all-86} the proposed heuristics were devised for
both minimizing communication costs and choosing an opportune site
for processing sub-queries among various network-connected database
systems. In both cases, the proposed heuristics were devised and
tuned for dealing with data distributed in several sites and their
application to other architectures might be neither viable nor
straightforward.

\section{Conclusions}\label{sec:conclusion}

In this paper we present some advanced techniques for the parallel instantiation
of ASP programs, and a parallel ASP instantiator based on the \dlv system.
In particular, we have proposed and implemented a three-level parallelization technique, dynamic load balancing, and granularity control strategies.
An experimental analysis outlines significant performance improvements, larger applicability w.r.t. existing approaches,
as well as nearly optimal efficiency and steady scalability of the implemented instantiator.

As far as future work is concerned, we are studying other techniques for further
exploiting parallelism in ASP systems, considering also the other phases of the computation.
Automatic determination of heuristics thresholds is also under investigation.

\section*{Acknowledgements}\label{sec:ack}
This work has been partially supported by Regione Calabria and EU under POR
Calabria FESR 2007-2013 within the PIA project of DLVSYSTEM s.r.l.

\bibliographystyle{acmtrans}

\appendix
\section{Splitting the Extension of a Literal}\label{app:split}
In Figure A1 is detailed an implementation of procedure {\em SplitExtension}, 
which plays a central role in the single-rule parallelization algorithms presented in Section~3.2.
In particular, function {\em SplitExtension} partitions the
extension of $l$ (stored in $S$ and $\DNF$) into $n$  splits.
In order to avoid useless copies, each split is virtually identified by means of
iterators over $S$ and $\DNF$, representing ranges of instances.

\begin{figure}[h!]
\begin{tabbing}\footnotesize
{\bf Procedure} {\em SplitExtension \/}(\= $l$: Literal; $n$:integer; $S$: SetOfAtoms; $\DNF$: SetOfAtoms, \\
\hspace*{1.2cm} {\bf var} vector$<$VirtualSplit$>$ $V$)\\
\hspace*{0.2cm} \= \kill
\hspace*{0.4cm} \= \hspace*{0.4cm} \= \hspace*{0.4cm} \=
\hspace*{0.4cm} \= \hspace*{0.4cm} \= \hspace*{3cm} \=\kill
{\bf begin} \\
\> integer $size$:=  $\lfloor$ ($S$.size() + \DNF.size())/ $n$ $\rfloor$; \\
\> integer $i$:= 0; \ \  AtomsIterator $it$ := $S$.begin(); \\
\>{\bf while} $i$ $<$ $\lfloor$$S$.size()/$size$$\rfloor$ {\bf do} \ \ // {\em possibly, build splits with atoms from $S$}\\
\>\> $V$[$i$].SetIterators\_S($it$, $it$+$size$); \ \ $it$ := $it$ + $size$; \ \ $i$=$i$+1;\\
\>{\bf end while}\\
\> {\bf if} $it$ $<$ $S$.end() {\bf then} \ \ // {\em possibly, build a split mixing $S$ and \DNF atoms} \\
\>\> $V$[$i$].SetIterators\_S($it$, $S$.end()); \ \ $it$ := \DNF.begin(); \\
\>\> integer $k$ := $size$ - Size($V[i]$); \\
\>\> {\bf if} \DNF.size() $<$ $k$ \\
\>\>\> $V$[$i$].SetIterators\_\DNF($it$, \DNF.end()); \ \  $it$ = \DNF.end();\\
\>\> {\bf else}\\
\>\>\> $V$[$i$].SetIterators\_\DNF($it$, $it$+$k$); \ \  $it$ = $it$+$k$; \ \ $i$ = $i$+1;\\
\>{\bf while} $i$ $<$ $\lfloor$($S$.size()+\DNF.size())/$size$$\rfloor$ {\bf do} \ \ // {\em possibly, build splits with atoms from \DNF}\\
\>\> $V$[$i$].SetIterators\_\DNF($it$, $it$+$size$); \ \ $it$ := $it$ + $size$; \ \ $i$=$i$+1;\\
\>{\bf end while}\\
\>{\bf if} $it$ $<$ \DNF.end() {\bf then}\\
\>\> $V$[$i$].SetIterators\_\DNF($it$, \DNF.end());\\
{\bf end Procedure} \\
\end{tabbing}\vspace{-0.3cm}
\caption{Splitting the extension of a literal.} \vspace{-0.2cm}
\end{figure}\label{fig:splitextension}

\noindent More in detail, for each  split, an instance of {\em VirtualSplit} is created containing two iterators over $S$
(resp. \DNF), namely $S\_begin$ and $S\_end$ (resp.
$\DNF\_begin$,$\DNF\_end$), indicating the instances of $l$ from $S$
(resp. \DNF) that belong to the split. 
The procedure starts by building splits with atoms from $S$; then, it proceeds
by considering atoms from \DNF. Note that, in general, a split may mix ground atoms 
from both $S$ and \DNF. 

\clearpage

\section{Detailed Results: 3Colorability and Hamiltonian Path}
\label{app:results}

Tables~\ref{tab_1} and~\ref{tab_2} contain detailed results for the benchmarks 3Colorability and Hamiltonian Path.
We recall that, in the case of 3Colorability, were generated $18$ simplex graphs by means of
the Stanford GraphBase library using the function $simplex(n,n,-2,0,$ $0,0,0)$, where $80\leq n \leq 250$;
whereas, for the Hamiltonian Path benchmark, $14$ graphs were generated by using
a tool by Patrik Simons, with $n$ nodes  with $1000 \leq n \leq 12000$.

In detail, Table~\ref{tab_1} reports the results of an experimental analysis aimed at comparing the
effects of the single rule parallelism with the first two levels.
The first column reports the problem considered, whereas the
next columns report the results for four version of the instantiator:
($i$) \serial where parallel techniques are not applied,
$(ii)$ \kali where components and rules parallelism are applied,
$(iii)$ \splitonly where only the single rule level is applied, and
$(iv)$ \paral in which all the three levels are applied.

Table~\ref{tab_2} reports the results of a scalability analysis on the instantiator \paral,
which exploits all the three parallelism levels. In particular, both the average instantiation times 
and the efficiencies are reported by considering the effects of increasing both the size of the instances 
and the number of available processors (from 1 up to 8 CPUs).%

\begin{table}[b!]
{
\begin{tabular}{| c | c | c | c | c |}
\cline{1-5}
Problem & \serial & \kali & \splitonly & \paral \\
\cline{1-5}
$3-Col_1$ & 8.61 (0.00) & 8.55 (0.09) & 1.27 (0.03) & 1.26 (0.03) \\
$3-Col_2$ & 14.16 (0.00) & 13.70 (0.04) & 1.90 (0.01) & 1.90 (0.01) \\
$3-Col_3$ & 20.44 (0.00) & 20.08 (0.10) & 2.85 (0.12) & 2.79 (0.03) \\
$3-Col_4$ & 30.92 (0.00) & 29.58 (0.17) & 4.09 (0.01) & 4.28 (0.30) \\
$3-Col_5$ & 42.13 (0.00) & 41.36 (0.06) & 5.62 (0.01) & 5.67 (0.13) \\
$3-Col_6$ & 59.38 (0.00) & 57.08 (0.24) & 8.03 (0.21) & 7.81 (0.03) \\
$3-Col_7$ & 78.64 (0.00) & 79.19 (0.07) & 10.29 (0.11) & 10.36 (0.15) \\
$3-Col_8$ & 104.45 (0.00) & 102.99 (0.34) & 13.70 (0.04) & 13.65 (0.04) \\
$3-Col_9$ & 133.72 (0.00) & 132.11 (1.96) & 17.58 (0.11) & 17.47 (0.08) \\
$3-Col_{10}$ & 177.28 (0.00) & 171.71 (0.84) & 23.00 (0.08) & 22.70 (0.37) \\
$3-Col_{11}$ & 220.07 (0.00) & 218.34 (0.75) & 29.40 (0.23) & 28.85 (0.05) \\
$3-Col_{12}$ & 281.29 (0.00) & 281.61 (0.65) & 37.22 (0.30) & 37.33 (0.28) \\
$3-Col_{13}$ & 347.97 (0.00) & 346.69 (1.65) & 45.48 (1.35) & 45.88 (1.12) \\
$3-Col_{14}$ & 420.88 (0.00) & 432.82 (2.02) & 59.86 (2.46) & 58.39 (0.74) \\
$3-Col_{15}$ & 528.20 (0.00) & 536.62 (1.00) & 72.39 (3.07) & 71.35 (0.44) \\
$3-Col_{16}$ & 673.24 (0.00) & 644.07 (2.99) & 85.75 (1.22) & 86.46 (0.63) \\
$3-Col_{17}$ & 786.43 (0.00) & 784.43 (1.00) & 103.16 (0.95) & 102.44 (3.54) \\
$3-Col_{18}$ & 965.36 (0.00) & 966.00 (2.31) & 120.09 (2.18) & 124.70 (1.08) \\
\cline{1-5}
$HP_1$ & 0.07 (0.00) & 0.05 (0.00) & 0.03 (0.00) & 0.03 (0.00) \\ 
$HP_2$ & 3.50 (0.00) & 3.21 (0.00) & 0.56 (0.01) & 0.57 (0.01) \\ 
$HP_3$ & 13.24 (0.00) & 12.42 (0.01) & 1.96 (0.01) & 1.97 (0.04) \\ 
$HP_4$ & 29.28 (0.00) & 27.85 (0.06) & 4.21 (0.10) & 4.18 (0.05) \\ 
$HP_5$ & 51.80 (0.00) & 49.28 (0.16) & 7.38 (0.13) & 7.33 (0.03) \\ 
$HP_6$ & 80.87 (0.00) & 77.24 (0.05) & 11.56 (0.07) & 11.56 (0.09) \\ 
$HP_7$ & 117.16 (0.00) & 112.35 (0.56) & 16.29 (0.02) & 16.77 (0.23) \\ 
$HP_8$ & 160.79 (0.00) & 154.58 (0.45) & 22.44 (0.15) & 22.45 (0.16) \\ 
$HP_{10}$ & 212.84 (0.00) & 205.60 (0.39) & 29.23 (0.07) & 29.57 (0.10) \\ 
$HP_{11}$ & 274.43 (0.00) & 263.61 (1.11) & 37.60 (0.24) & 37.14 (0.29) \\ 
$HP_{12}$ & 343.05 (0.00) & 331.16 (0.50) & 46.63 (0.09) & 47.68 (0.03) \\ 
$HP_{13}$ & 422.72 (0.00) & 406.36 (0.64) & 57.54 (0.48) & 57.66 (1.05) \\ 
$HP_{14}$ & 510.15 (0.00) & 492.28 (1.68) & 68.95 (0.56) & 70.26 (0.38) \\
\cline{1-5}
\end{tabular}
\caption{3Colorability and Hamiltonian Path results: Average instantiation times in seconds (standard deviation) }
\label{tab_1}
}
\end{table}

\begin{landscape}
\scriptsize
 
\begin{tabular}{@{\extracolsep{\fill}}|c|r|r|r|r|r|r|r|r||r|r|r|r|r|r|r|}
\cline{1-16}

Problem & serial & 2 proc  & 3 proc & 4 proc & 5 proc & 6 proc & 7 proc & 8 proc &  2 proc  & 3 proc & 4 proc & 5 proc & 6 proc & 7 proc & 8 proc\\

\cline{1-16}
$3-Col_1$ & 8.61 (0.00) &  4.41 (0.02) &  3.02 (0.02) &  2.29 (0.03) &  1.86 (0.03) &  1.59 (0.02) &  1.38 (0.03) &  1.26 (0.03) &  0.98 & 0.95 & 0.94 & 0.93 & 0.90 & 0.89 & 0.85 \\
$3-Col_2$ & 14.16 (0.00) &  7.15 (0.02) &  4.83 (0.01) &  3.65 (0.03) &  2.93 (0.00) &  2.51 (0.04) &  2.19 (0.03) &  1.90 (0.01) & 0.99 & 0.98 & 0.97 & 0.97 & 0.94 & 0.92 & 0.93 \\
$3-Col_3$ & 20.44 (0.00) &  10.49 (0.01) &  7.11 (0.06) &  5.35 (0.02) &  4.31 (0.03) &  3.68 (0.11) &  3.12 (0.04) &  2.79 (0.03) & 0.97 & 0.96 & 0.96 & 0.95 & 0.93 & 0.94 & 0.92 \\
$3-Col_4$ & 30.92 (0.00) &  15.74 (0.27) &  10.78 (0.19) &  7.97 (0.02) &  6.35 (0.02) &  5.37 (0.03) &  4.77 (0.20) &  4.28 (0.30) & 0.98 & 0.96 & 0.97 & 0.97 & 0.96 & 0.93 & 0.90 \\
$3-Col_5$ & 42.13 (0.00) &  21.76 (0.43) &  14.58 (0.20) &  10.93 (0.21) &  8.86 (0.12) &  7.46 (0.15) &  6.57 (0.17) &  5.67 (0.13) & 0.97 & 0.96 & 0.96 & 0.95 & 0.94 & 0.92 & 0.93 \\
$3-Col_6$ & 59.38 (0.00) &  29.96 (0.05) &  19.94 (0.00) &  15.08 (0.07) &  12.16 (0.11) &  10.15 (0.07) &  8.87 (0.14) &  7.81 (0.03) & 0.99 & 0.99 & 0.98 & 0.98 & 0.98 & 0.96 & 0.95 \\
$3-Col_7$ & 78.64 (0.00) &  40.25 (0.19) &  26.66 (0.42) &  20.31 (0.19) &  16.24 (0.34) &  13.44 (0.13) &  11.56 (0.10) &  10.36 (0.15) & 0.98 & 0.98 & 0.97 & 0.97 & 0.98 & 0.97 & 0.95 \\
$3-Col_8$ & 104.45 (0.00) &  53.79 (0.33) &  35.70 (0.35) &  26.85 (0.36) &  21.45 (0.28) &  17.96 (0.18) &  15.55 (0.14) &  13.65 (0.04) & 0.97 & 0.98 & 0.97 & 0.97 & 0.97 & 0.96 & 0.96 \\
$3-Col_9$ & 133.72 (0.00) &  68.53 (0.22) &  46.32 (0.17) &  34.79 (0.34) &  27.60 (0.18) &  23.34 (0.13) &  20.18 (0.19) &  17.47 (0.08) & 0.98 & 0.96 & 0.96 & 0.97 & 0.95 & 0.95 & 0.96 \\
$3-Col_{10}$ & 177.28 (0.00) &  89.29 (0.65) &  60.65 (0.58) &  44.91 (0.24) &  36.34 (0.26) &  30.54 (0.51) &  26.35 (0.06) &  22.70 (0.37) & 0.99 & 0.97 & 0.99 & 0.98 & 0.97 & 0.96 & 0.98 \\
$3-Col_{11}$ & 220.07 (0.00) &  113.03 (1.76) &  76.25 (0.54) &  57.71 (0.92) &  45.77 (0.52) &  38.94 (0.44) &  32.97 (0.37) &  28.85 (0.05) & 0.97 & 0.96 & 0.95 & 0.96 & 0.94 & 0.95 & 0.95 \\
$3-Col_{12}$ & 281.29 (0.00) &  143.00 (2.39) &  98.02 (0.38) &  73.68 (0.84) &  59.08 (0.27) &  49.56 (0.72) &  42.73 (0.17) &  37.33 (0.28) & 0.98 & 0.96 & 0.95 & 0.95 & 0.95 & 0.94 & 0.94 \\
$3-Col_{13}$ & 347.97 (0.00) &  178.98 (4.07) &  123.22 (2.62) &  90.09 (1.79) &  71.84 (1.48) &  61.42 (2.13) &  53.32 (0.52) &  45.88 (1.12) & 0.97 & 0.94 & 0.97 & 0.97 & 0.94 & 0.93 & 0.95 \\
$3-Col_{14}$ & 420.88 (0.00) &  224.85 (3.82) &  152.87 (2.07) &  113.88 (0.92) &  92.51 (1.66) &  76.46 (0.57) &  66.68 (0.40) &  58.39 (0.74) & 0.94 & 0.92 & 0.92 & 0.91 & 0.92 & 0.90 & 0.90 \\
$3-Col_{15}$ & 528.20 (0.00) &  278.26 (5.67) &  184.32 (1.48) &  138.84 (1.63) &  111.29 (2.58) &  94.25 (0.97) &  80.58 (0.24) &  71.35 (0.44) & 0.95 & 0.96 & 0.95 & 0.95 & 0.93 & 0.94 & 0.93 \\
$3-Col_{16}$ & 673.24 (0.00) &  331.55 (4.12) &  224.66 (4.57) &  169.90 (7.72) &  137.37 (1.34) &  111.57 (2.10) &  97.50 (0.56) &  86.46 (0.63) & 1.02 & 1.00 & 0.99 & 0.98 & 1.01 & 0.99 & 0.97 \\
$3-Col_{17}$ & 786.43 (0.00) &  403.02 (5.80) &  263.37 (2.69) &  207.01 (5.21) &  158.80 (5.81) &  135.89 (1.64) &  116.68 (3.89) &  102.44 (3.54) & 0.98 & 1.00 & 0.95 & 0.99 & 0.96 & 0.96 & 0.96 \\
$3-Col_{18}$ & 965.36 (0.00) &  473.35 (4.88) &  312.04 (6.90) &  238.99 (4.15) &  190.61 (1.65) &  159.41 (4.64) &  140.46 (1.49) &  124.70 (1.08) & 1.02 & 1.03 & 1.01 & 1.01 & 1.01 & 0.98 & 0.97 \\

\cline{1-16}
$HP_1$&3.50 (0.00) &  1.82 (0.01) &  1.27 (0.00) &  1.00 (0.02) &  0.80 (0.00) &  0.70 (0.00) &  0.63 (0.00) &  0.57 (0.01) &  0.96 & 0.92 & 0.88 & 0.88 & 0.83 & 0.79 & 0.77 \\
$HP_2$&13.24 (0.00) &  6.84 (0.05) &  4.64 (0.01) &  3.59 (0.05) &  2.90 (0.04) &  2.48 (0.01) &  2.19 (0.06) &  1.97 (0.04) & 0.97 & 0.95 & 0.92 & 0.91 & 0.89 & 0.86 & 0.84 \\
$HP_3$&29.28 (0.00) &  15.63 (0.28) &  10.41 (0.26) &  7.90 (0.19) &  6.30 (0.02) &  5.52 (0.22) &  4.67 (0.03) &  4.18 (0.05) & 0.94 & 0.94 & 0.93 & 0.93 & 0.88 & 0.90 & 0.88 \\
$HP_4$&51.80 (0.00) &  26.55 (0.03) &  18.05 (0.02) &  13.82 (0.16) &  11.14 (0.04) &  9.52 (0.14) &  8.22 (0.05) &  7.33 (0.03) & 0.98 & 0.96 & 0.94 & 0.93 & 0.91 & 0.90 & 0.88 \\
$HP_5$&80.87 (0.00) &  42.60 (0.18) &  28.68 (0.38) &  21.61 (0.05) &  17.56 (0.12) &  15.07 (0.13) &  13.05 (0.01) &  11.56 (0.09) & 0.95 & 0.94 & 0.94 & 0.92 & 0.89 & 0.89 & 0.87 \\
$HP_6$&212.84 (0.00) &  110.90 (0.48) &  74.50 (0.56) &  56.16 (0.26) &  45.62 (0.35) &  38.10 (0.42) &  33.46 (0.11) &  29.57 (0.10) &0.96 & 0.95 & 0.93 & 0.92 & 0.91 & 0.89 & 0.87 \\
$HP_7$&274.43 (0.00) &  141.00 (0.59) &  94.69 (0.84) &  72.07 (0.11) &  58.30 (0.35) &  48.86 (0.15) &  42.67 (0.21) &  37.14 (0.29) & 0.97 & 0.96 & 0.94 & 0.94 & 0.92 & 0.90 & 0.90 \\
$HP_8$&160.79 (0.00) &  82.52 (0.35) &  56.06 (0.23) &  42.64 (0.34) &  34.09 (0.11) &  29.26 (0.05) &  25.50 (0.32) &  22.45 (0.16) & 0.96 & 0.95 & 0.95 & 0.93 & 0.93 & 0.91 & 0.90 \\
$HP_9$&343.05 (0.00) &  176.40 (0.89) &  118.84 (0.47) &  89.88 (0.53) &  73.53 (0.19) &  61.61 (0.43) &  53.33 (0.10) &  47.68 (0.03) & 0.97 & 0.97 & 0.95 & 0.94 & 0.94 & 0.92 & 0.92 \\
$HP_{10}$&117.16 (0.00) &  60.84 (0.29) &  41.05 (0.49) &  31.40 (0.32) &  25.55 (0.13) &  21.43 (0.17) &  18.81 (0.08) &  16.77 (0.23) & 0.97 & 0.96 & 0.95 & 0.93 & 0.93 & 0.92 & 0.90 \\
$HP_{11}$&422.72 (0.00) &  218.38 (0.51) &  146.26 (0.38) &  110.34 (0.56) &  89.77 (0.19) &  75.15 (0.25) &  66.01 (0.59) &  57.66 (1.05) & 0.97 & 0.96 & 0.96 & 0.94 & 0.94 & 0.91 & 0.92 \\
$HP_{12}$&510.15 (0.00) &  261.06 (2.47) &  173.57 (1.93) &  132.88 (0.21) &  107.51 (0.63) &  90.66 (0.80) &  78.44 (0.44) &  70.26 (0.38) & 0.98 & 0.98 & 0.96 & 0.95 & 0.94 & 0.93 & 0.91 \\
\cline{1-16}

\end{tabular}

 \captionof{table}{3Colorability and Hamiltonian Path results: average instantiation times in seconds (standard deviation), efficiency}
  \label{tab_2}
\end{landscape}

\section{Application of the Single Rule Parallelism}
\label{AppSingleRule}

In the following is reported a complete example of the application of the single rule parallelism for computing
in parallel the instantiation of a rule.
Consider the following program \p encoding the 3-Colorability problem:

\begin{small}
\begin{dlvcode}
(r)\ \ \ col(X,red)\ \Or\ col(X,yellow)\ \Or\ col(X,green)\ \derives\ node(X). \\
(c)\ \ \ \derives\ edge(X,Y), col(X,C),\ col(Y,C).
\end{dlvcode}
\end{small}

Assume that, after the instantiation of rule $r$, 
the extensions of predicate $node$ and predicate $col$, are the ones reported in Table~\ref{ext_1}, and 
that the extension of the predicate $edge$ is the one reported in Table~\ref{ext_2}.

\begin{table}[h!]
 \begin{tabular}{|c|c|}
  \cline{1-2}
  & Predicates extension \\
  \cline{1-2}
1&$node(a)$ \\
2&$node(b)$ \\
3&$node(c)$ \\
4&$node(d)$ \\
\cline{1-2}
\cline{1-2}
1 & $col(a,red)$ \\
2 & $col(a,yellow)$ \\ 
3 & $col(a,green)$ \\
4 & $col(b,red)$ \\
5 & $col(b,yellow)$ \\
6 & $col(b,green)$ \\
7 & $col(c,red)$ \\
8 & $col(c,yellow)$ \\
9 & $col(c,green)$\\
10 & $col(d,red)$\\
11 & $col(d,yellow)$\\
12 & $col(d,green)$ \\
  \cline{1-2}
 \end{tabular} 
\caption{Extension of the predicate $node$ and $col$.}
\label{ext_1}
\end{table}

\noindent Suppose now that the heuristics suggests to perform the single rule level of parallelism
for the instantiation of the constraint $(c)$, and suppose that the extension of predicate $edge$ is split in two.
Then, the extension of the predicate $edge$ is partitioned into two subsets which appear divided by an horizontal line in Table~\ref{ext_2}.
The instantiation of constraint $(c)$ is carried out in parallel by two separate processes, 
say $p_1$, and $p_2$, which will consider as extension of $edge$, respectively, the two splits depicted in Table~\ref{ext_2}.
Process $p_1$ produces the following ground constraints:
\begin{small}
\begin{dlvcode}
\derives edge(a,b), col(a,red),\ col(b,red). \\ 
\derives edge(a,b), col(a,yellow),\ col(b,yellow). \\ 
\derives edge(a,b), col(a,green),\ col(b,green). \\ 
\derives edge(b,c), col(a,red),\ col(b,red). \\
\derives edge(b,c), col(a,yellow),\ col(b,yellow). \\
\derives edge(b,c), col(a,green),\ col(b,green). 
\end{dlvcode}
\end{small}
\noindent whereas, process $p_2$ produces the following ground constraints:
\begin{small}
\begin{dlvcode}
\derives edge(b,d), col(b,red),\ col(d,red). \\ 
\derives edge(b,d), col(b,yellow),\ col(d,yellow). \\ 
\derives edge(b,d), col(b,green),\ col(d,green). \\ 
\derives edge(c,d), col(c,red),\ col(d,red). \\
\derives edge(c,d), col(c,yellow),\ col(d,yellow). \\
\derives edge(c,d), col(c,green),\ col(d,green). 
\end{dlvcode}
\end{small}

Tecnically speaking, this is obtained by a call to procedure {\em SplitExtension \/} described in~\ref{app:split}.
The procedure will create two {\em virtual splits}, say $V_1$ and $V_2$, with:
\begin{itemize}
\item[] $V_1.S_{begin} = edge(a,b)$ \ \ \ \ $V_1.\DNF_{begin} = \bot$
\item[] $V_1.S_{end} = edge(b,c)$  \ \ \ \ \ \ \ $V_1.\DNF_{end} = \bot $
\item[]$V_2.S_{begin} = edge(b,d)$ \ \ \ \ $V_2.\DNF_{begin} = \bot$
\item[] $V_2.S_{end} = edge(c,d)$. \ \ \ \ \ $V_2.\DNF_{end} = \bot$
\end{itemize}

\noindent where $\bot$ indicates a null iterator 
(usually indicating an iterator that has moved after the end of a container), which, in this case, it is used to represent that  that no split is created containing instances from $\DNF$.

\begin{table}[t!]
 \begin{tabular}{|c|c|}
  \cline{1-2}
  & Predicate extension \\
  \cline{1-2}
    1 & $edge(a,b)$ \\
    2 & $edge(b,c)$ \\
  \cline{1-2}
    3 & $edge(b,d)$ \\
    4 & $edge(c,d)$ \\
  \cline{1-2}
 \end{tabular}
\caption{Extension of the predicate $edge$.}
\label{ext_2}
\end{table}

\end{document}